\documentclass[usenatbib]{mn2e}
\usepackage{graphicx}
\newcommand{\text}[1]{\quad\mbox{#1}\quad}
\newcommand{\spr}[2]{\bmath{#1} \!\cdot\! \bmath{#2}}
\newcommand{\vpr}[2]{\bmath{#1} \!\times\! \bmath{#2}}

\newcommand{\vgrad}[1]{\nabla{#1}}

\newcommand{\pder}[2]{\frac{\partial #1}{\partial #2}}
\newcommand{\Pd}[1]{\partial_{#1}}

\newcommand{\thetak}{\vartheta_k}
\newcommand{\thetag}{\vartheta_g}
\newcommand{\massint}{k}
\newcommand{\enint}{\mu}
\newcommand{\aj}{AJ}
\newcommand{\apj}{ApJ}
\newcommand{\apjs}{ApJS}

\newcommand{\mnras}{MNRAS}

\newcommand{\aap}{A\&A}

\title{Magnetic acceleration of relativistic AGN jets} 
\author[S.~S. Komissarov et al.]{Serguei
S.~Komissarov,$^{1}$\thanks{E-Mail: serguei@maths.leeds.ac.uk (SSK);
bmv@maths.leeds.ac.uk (MVB); vlahakis@phys.uoa.gr (NV);
arieh@jets.uchicago.edu (AK)} Maxim V.~Barkov,$^{1,2}$\footnotemark[1]
Nektarios Vlahakis$^{3}$\footnotemark[1] and Arieh
K\"onigl$^{4}$\footnotemark[1]\\ 
$^{1}$Department of Applied Mathematics, The University of Leeds,
Leeds, LS2 9GT\\
$^{2}$Space Research Institute, 84/32 Profsoyuznaya Street, Moscow
117997, Russia\\
$^{3}$Section of Astrophysics, Astronomy and Mechanics, 
Physics Department, University of Athens, 15784 Zografos, Athens, Greece\\
$^{4}$Department of Astronomy and Astrophysics and
Enrico Fermi Institute, University of Chicago, 5640 South Ellis
Avenue,\\ \hskip 1cm Chicago, IL 60637, USA}
\begin{document}
\date{Received/Accepted}
\maketitle
                                                                              
\begin{abstract}

We present numerical simulations of axisymmetric, magnetically driven
relativistic jets. Our special-relativistic, ideal-MHD numerical scheme
is specifically designed to optimize accuracy and resolution and to
minimize numerical dissipation. In addition, we implement a
grid-extension method that reduces the computation time by up to three
orders of magnitude and makes it possible to follow the flow up to six
decades in spatial scale.  To eliminate the dissipative effects induced
by a free boundary with an ambient medium we assume that the flow is
confined by a rigid wall of a prescribed shape, which we take to be
$z\propto r^a$ (in cylindrical coordinates, with $a$ ranging from 1 to
3). We also prescribe, through the rotation profile at the inlet
boundary, the injected poloidal current distribution: we explore cases
where the return current flows either within the volume of the jet or on
the outer boundary.  The outflows are initially cold, sub-Alfv\'enic and
Poynting flux-dominated, with a total--to--rest-mass energy flux ratio
$\mu \sim 15$.  We find that in all cases they converge to a steady
state characterized by a spatially extended acceleration region. The
acceleration process is very efficient: on the outermost scale of the
simulation as much as $\sim 77\%$ of the Poynting flux has been
converted into kinetic energy flux, and the terminal Lorentz factor
approaches its maximum possible value ($\Gamma_\infty \simeq \mu$).  We
also find a high collimation efficiency: all our simulated jets 
(including the limiting case of an unconfined flow)
develop a cylindrical core. We argue that this could be the rule for
current-carrying outflows that start with a low initial Lorentz factor
($\Gamma_0 \sim 1$). Our conclusions on the high acceleration and
collimation efficiencies are not sensitive to the particular shape of
the confining boundary or to the details of the injected current
distribution, and they are qualitatively consistent with the
semi-analytic self-similar solutions derived by Vlahakis \& K\"onigl.
We apply our results to the interpretation of relativistic jets in AGNs:
we argue that they naturally account for the spatially extended
accelerations inferred in these sources ($\Gamma_\infty \ga 10$ attained
on radial scales $R\ga 10^{17}\, {\rm cm}$) and are consistent with the
transition to the matter-dominated regime occurring already at $R\ga
10^{16}\, {\rm cm}$.

\end{abstract}

\begin{keywords}
galaxies: active -- galaxies: jets -- methods: numerical -- MHD --
relativity
\end{keywords}
                                                                          
\section{Introduction}
\label{introduction}

There is strong evidence for relativistic motions in jets that emanate
from active galactic nuclei (AGNs). In particular, apparent
superluminal speeds $\beta_{\rm app}$ (in units of the speed of
light~$c$) as high as $\sim 40$ have been measured for radio
components on (projected) scales of $\sim 1-10\;$pc in the blazar
class of sources \citep[e.g.][]{J01}. \citet{J05} used a method based
on a comparison between the time scale of flux density decline and the
light-travel time across the imaged emission region to relate
$\beta_{\rm app}$ to the bulk Lorentz factor $\Gamma$ of the outflow;
they inferred that the Lorentz factors of blazar jets lie in the range
$\sim 5-40$, with the majority of quasar components having $\Gamma
\sim 16-18$ and with BL Lac objects possessing a more uniform $\Gamma$
distribution. \citet{C07} recently reached a similar conclusion on the
basis of probability arguments, inferring that roughly half the
sources in a flux density-limited, beamed sample have a value of
$\Gamma$ close to the measured $\beta_{\rm app}$. They further deduced
that the maximum Lorentz factor in their sample of 119 AGN jets is
$\sim 32$, close to the value of $\sim 40$ inferred for the jets
observed by \citet{J01,J05}.

The presence of relativistic bulk motions in blazar jets has been
independently indicated by measurements of rapid variations in the total
and polarized fluxes \citep[e.g.][]{Ha01,R06,B06,V06}. There is also
evidence that the relativistic speeds persist to large scales. For
example, apparent superluminal component motions have been measured in
the 3C 120 jet out to projected distances from the source of at least
$150\;$pc \citep{W01}, and it has been argued that the spectral
properties of the heads of extended (up to several hundred kiloparsecs)
jets can be explained in the context of a relativistic flow that is
decelerated to subrelativistic speeds at the termination shock that
advances into the ambient medium \citep{GK03}.

The main source of power of AGN jets is the rotational energy of the
central supermassive black hole \citep[e.g.][]{L76,BZ77} and/or its
accretion disk \citep[e.g.][]{B76}. The naturally occurring low mass
density and hence high magnetization of black-hole magnetospheres
suggests that the relativistic jets originate directly from the
black-hole ergosphere, whereas the disk surface launches a slower,
possibly nonrelativistic wind that surrounds and confines the highly
relativistic flow. This picture finds support in recent numerical
simulations \citep[e.g][]{McK04,DeV05}.  However, this issue is far from
settled and one cannot rule out the inner part of accretion disk as a
base of a relativistic outflow \citep[e.g.][]{VK04}.
The theory of relativistic, magnetically driven jets from black holes 
(and neutron stars) predicts highly magnetized flows, with the Poynting flux 
dominating the total energy output. At the jet emission site 
this energy has to be transferred to particles. This transfer 
may have the form of magnetic dissipation \citep[e.g.][]{B02,LB03} but the 
still commonly held view is that the Poynting energy is first converted into
bulk kinetic energy 
and only subsequently channeled into radiation
through shocks and other dissipative waves 
\citep[e.g.][]{BR74,BBR84}. The 
jet radiative efficiency is still one of the key debatable issues in the
theory of Poynting-dominated outflows.
In principle, slow magnetic 
dissipation in an expanding jet may also facilitate the gradual conversion of 
Poynting flux into bulk kinetic energy 
\citep{DS02}.        

When the inertia of the plasma 
is negligibly small its dynamics 
is well described by the approximation of force-free electrodynamics 
\citep[or magnetodynamics; e.g.][]{K02,KBL07}. The equations of
magnetodynamics are much simpler than those of magnetohydrodynamics (MHD),
which is what prompted their application to the study of the magnetic
acceleration of relativistic jets \citep[e.g.][]{B76,B02,N07}.
The solutions of these equations describe trans-Alfv\'enic flows whose
drift velocity approaches the speed of 
light at infinity, the location of the fast critical point in these models.
However, within this framework it is impossible to account for the
conversion of Poynting flux into plasma kinetic energy and to study
the issue of the acceleration efficiency efficiency.

The next simplest approximation that can be used to address the issue of
Poynting-to-kinetic energy conversion is ideal MHD.  
In this case one can obtain
exact semi-analytic solutions, although, because of the complexity of
the problem, this can only be done when the system possesses a
high degree of symmetry. 
This approach was pioneered by \citet{BP82}, who constructed
nonrelativistic semi-analytic solutions  for steady-state, cold,
self-similar (in the spherical radial coordinate) disk outflows.  
These solutions were generalized to the  
(special) relativistic MHD regime by \citet*{LCB92} and by \citet{C94}. 
They were further
investigated by \citet{VK03a,VK03b}, who also considered the effect of
thermal forces during the early phases of the
acceleration.\footnote{\citet{VK03a} focused on flows whose initial
poloidal velocity component is sub-Alfv\'enic, corresponding to the
poloidal magnetic field component dominating the azimuthal field
component at the top of the disk, whereas \citet{VK03b} discussed the
super-Alfv\'enic case in which the azimuthal component is dominant at
the base of the flow. In this paper we only consider outflows of the
first type.}  Solutions with similar properties were derived in
\citet{BN06} by linearizing about a force-free solution for a
paraboloidal field geometry.

A key property of the relativistic solutions derived in the
aforementioned studies is the extended nature of the acceleration
region: the bulk of the (poloidal) acceleration is effected by
magnetic pressure gradients (associated with the azimuthal magnetic
field component) and takes place beyond the classical
fast-magnetosonic point (a singular point of the Bernoulli equation).
\citet{VK03a} interpreted this behaviour (which was dubbed the
``magnetic nozzle'' effect by \citealt{LCB92}; see also \citealt{C89})
in terms of the distinction between the classical and the modified
fast-magnetosonic surfaces \citep[e.g.][]{B97}.  They pointed out that
the latter surface, which is the locus of the fast-magnetosonic
singular points of the combined Bernoulli and trans-field (or
Grad-Shafranov) equations, is the true causality surface (or ``event
horizon'') for the propagation of fast waves when the shape of the
field lines is obtained from the solution of the trans-field equation
(with the classical surface playing this role only when the shape of
the flux surfaces is predetermined). They argued that, in this case,
the acceleration continues all the way to (and possibly even past) the
modified fast-magnetosonic surface, which can lie well beyond the
classical one.\footnote{In the radially self-similar solutions
presented in \cite{VK03a}, the modified fast-magnetosonic surface
formally lies at an infinite distance from the origin.}  Another
general property of the cold MHD solutions is that, in the
current-carrying regime (where the poloidal components of the current
density and the magnetic field are antiparallel) they collimate
(asymptotically) to cylinders. Furthermore, the asymptotic Lorentz
factor corresponds to a rough equipartition between the Poynting and
kinetic energy fluxes.

The continuation of the acceleration process beyond the classical
fast-magnetosonic surface is evidently a general characteristic of
steady-state MHD solutions that applies also to nonrelativistic jets
\citep[e.g.][]{V00}.  This behaviour should, however, be more clearly
discerned in observations of relativistic flows, where the proper
speed $\Gamma \beta$ can increase by a large factor between the
classical and the modified singular surfaces.  In contrast, the
magnetic acceleration of non-relativistic flows is almost complete at
the classical fast point.  This striking difference has a very simple
origin. For nonrelativistic flows the criticality condition at the
classical fast-magnetosonic point implies equipartition between the
magnetic energy and the kinetic energy of poloidal motion. The kinetic
energy can therefore increase by at most a factor of 2 beyond this
point. However, relativistic flows remain magnetically dominated at
the fast-magnetosonic point, which means that there is an ample
remaining supply of magnetic energy that can be used for flow
acceleration downstream of this point (e.g. \citealp{K04}).

In the case of AGNs there have indeed been indications from a growing
body of data that the associated relativistic jets undergo the bulk of
their acceleration on scales that are of the order of those probed by
very-long-baseline radio interferometry. In one line of reasoning, the
absence of bulk-Comptonization spectral signatures in blazars has been
used to infer that jet Lorentz factors $\ga 10$ are only attained on
scales $\ga 10^{17}\ {\rm cm}$ \citep{S05}. There have also been
explicit inferences of component acceleration based on radio proper
motion and X-ray emission measurements for the jets in the quasars 3C
345 \citep{U97} and 3C 279 \citep{P03}. Extended acceleration in the
3C 345 jet has been independently indicated by the higher apparent
speeds of jet components located further away from the nucleus
\citep{LR05} and by the observed luminosity variations of the moving
components \citep{LZ99}. Similar effects in other blazars
\citep[e.g.][]{H01} suggest that parsec-scale acceleration to
relativistic speeds may be a common feature of AGN jets. \citet{VK04}
argued that these observations are most naturally interpreted in terms
of magnetic driving and employed self-similar relativistic jet
solutions to generate model fits to the 3C 345 data in support of this
conclusion.

While the semi-analytic solutions have been useful in indicating basic
properties of the magnetic acceleration process and in providing
valuable clues to the interpretation of the observational data, more
general solutions are needed to confirm these results and to gain a
fuller understanding of the generation of relativistic jets in
AGNs. In particular, numerical simulations are needed to find out
whether the self-similar model captures the essential properties of
outflows that obey realistic boundary conditions and that are not
required to be in a steady state. Among the questions that such
simulations could answer are: (1) Do disk outflows in fact approach a
steady state, and, if they do, is that state stable?  (2) Is the
acceleration indeed generally extended, and to what extent does the
asymptotic state of the self-similar solutions approximate the
far-field behaviour of more realistic outflows? (3) Do any new traits
emerge when the restrictions imposed by the self-similarity assumption
are removed? Of particular interest is the question of the ability of
the magnetic driving mechanism to accelerate outflows to high Lorentz
factors with high efficiency over astrophysically relevant distance
scales.  Another important question is whether highly relativistic
flows can be strongly collimated by purely magnetic stresses.  There
have been lingering doubts over these issues in the literature (see
Section~\ref{theory}), and although they have already received
tentative answers, a full numerical study could help to settle them
once and for all.

Although there have already been several reported simulations of the
formation of jets in black-hole accretion flows using relativistic (in
fact, {\it general}-relativistic) MHD codes, so far they have provided
only partial answers to the above questions. The existing calculations
indicate that magnetic acceleration indeed operates over several decades
in radius and can accelerate jets to relativistic speeds. However, the
extended nature of the acceleration typically results in the bulk
Lorentz factor reaching only a small fraction of its potential
asymptotic value by the time the simulation is terminated. 
For example, in the longest 
jet simulated to date, which extended to $\sim 10^4$ times the gravitational 
radius $r_g$ of the central black hole \citep{McK06}, 
the Lorentz factor on the largest computed scale was $\sim 10$, which is just 
$\sim 10^{-2}-10^{-3}$ of the estimated asymptotic value. This impressive 
simulation deals with an extremely complex system of which the jet is only one 
component, the other being the black hole, the accretion disk, the disk
corona, the low-speed  ``wall jet,'' and their surroundings. Ultimately,
this is the kind of simulation 
one wants to carry out in order to fully understand the dynamics of AGN
outflows. However, they are 
also very challenging from the computational point of view. One major
concern ins that, in view of the extended 
nature of its acceleration, the jet is particularly vulnerable to numerical 
diffusion and dissipation. These numerical effects may 
partly explain why the quantity $\Gamma_\infty$ in the above-cited paper, which 
is the same as our $\enint$ (equation~\ref{kap-def}) and should be a field-line
constant, in fact decreases by about one 
order of magnitude along a mid-level field line in the simulated jet
(see Fig.~7 in \citealt{McK06}).

In this paper we address the above questions through numerical
simulations specifically designed for investigating the key aspects of
ideal-MHD acceleration of relativistic jets.  In the first place, we
use a numerical scheme based on a linear Riemann solver \citep{K99} that
does not need a large artificial diffusion for numerical stability. This
distinguishes it from most other schemes for relativistic MHD, including
those that are based on HLL, KT and similar flux prescriptions
\citep[e.g.][]{GMT,Due,KSK,Ann,SS,AHLN,Del,Ant}.  Simple one-dimensional
tests suggest that this should lead to a noticeably greater accuracy in
two-dimensional problems that involve stationary flows that are aligned
with the computational grid \citep{K06}.  Secondly, instead of studying
jet propagation through some ambient medium, we consider the case of a
flow in a funnel with solid walls. This allows us to avoid the errors
that would otherwise be caused by numerical mass diffusion and dissipation 
at the interface.  Finally, we employ elliptical (or spherical) coordinates
adapted to the chosen paraboloidal (or conical) shape of the funnel.
This allows us to have the jet well resolved everywhere (using a fixed
number of grid points across the funnel) and to benefit from the close
alignment of the flow with the computational grid.  These careful
measures in conjunction with a grid-extension method enable us, for the
first time, to track the acceleration and collimation processes to their
completion.

We describe the basic equations in Section~\ref{equations} and the
numerical calculations in Section~\ref{simulations}. The simulation
results are presented in Section~\ref{results} and discussed in
Section~\ref{discussion}. We summarize in Section~\ref{conclusion}.

\section{Basic equations}
\label{equations}

Since most of the acceleration takes place far away from the source,
we assume that the space-time is flat. Moreover, the flow is described
in an inertial frame at rest relative to the source. In this case we
can write the system of ideal relativistic MHD as follows. The {\it
continuity equation}
\begin{equation}
(1/c)\Pd{t}(\sqrt{-g}\rho u^t)+ \Pd{i}(\sqrt{-g}\rho u^i)=0\, ,
\label{cont1}
\end{equation}
where $\rho$ is the rest mass density of matter, $u^\nu$ is its
4-velocity, and $g$ is the determinant of the metric tensor; the {\it
energy-momentum equations}
\begin{equation} 
(1/c)\Pd{t}(\sqrt{-g}T^t_{\ \nu})+ \Pd{i}(\sqrt{-g}T^i_{\ \nu})=
\frac{\sqrt{-g}}{2} \Pd{\nu}(g_{\alpha\beta}) T^{\alpha\beta}\, ,
\label{en-mom1} 
\end{equation} 
where $T^{\kappa\nu}$ is the total stress-energy-momentum tensor; the
{\it induction equation}
\begin{equation} (1/c)\Pd{t}(B^i)+e^{ijk}\Pd{j}(E_k) =0\, , \label{ind1}
\end{equation} 
where $e_{ijk} = \sqrt{\gamma} \epsilon_{ijk} $ is the Levi-Civita
tensor of the absolute space ($\epsilon_{123}=1$ for right-handed
systems and $\epsilon_{123}=-1$ for left-handed ones) and $\gamma$ is
the determinant of the spatial part of the metric tensor
($\gamma_{ij}=g_{ij}$); the {\it solenoidal condition}
\begin{equation}
\Pd{i}(\sqrt{\gamma} B^i) =0\, .
\end{equation}

The total stress-energy-momentum tensor, $T^{\kappa\nu}$, is a sum of
the stress-energy momentum tensor of matter
\begin{equation}
T_{(m)}^{\kappa\nu} = wu^\kappa u^\nu /c^2 + p g^{\kappa\nu}\, ,
\end{equation}
where $p$ is the thermodynamic pressure and $w$ is the enthalpy per
unit volume, and the stress-energy momentum tensor of the
electromagnetic field
\begin{equation}
   T_{(e)}^{\kappa\nu} = \frac{1}{4\pi}\left[F^{\kappa\alpha} F^\nu_{\
   \alpha} - \frac{1}{4}(F^{\alpha\beta}F_{\alpha\beta})g^{\kappa\nu}
   \right]\, ,
\end{equation}
where $F^{\nu\kappa}$ is the Maxwell tensor of the electromagnetic
field.  The electric and magnetic field are defined as measured by an
observer stationary relative to the spatial grid, which gives
\begin{equation}
     B^i= \frac{1}{2}e^{ijk} F_{jk}
\label{B-def}
\end{equation}
and
\begin{equation}
   E_i = F_{it}\, .
\label{E-def}
\end{equation}

In the limit of ideal MHD
\begin{equation}
  E_i=-e_{ijk}v^jB^k/c\, ,
\label{perf-cond}
\end{equation}
where $v^i=u^i/u^t$ is the usual 3-velocity of the plasma.
 
We use an isentropic equation of state
\begin{equation}
  p=Q\rho^s\, ,
\label{eos}
\end{equation}
where $Q=$const and $s=4/3$. Since we are interested in the magnetic
acceleration of cold flows, we make $Q$ very small, so the gas pressure
is never a dynamical factor. 
This relation enables us to exclude the
energy equation from the integrated 
system.  However, the momentum equation remains intact, including the
nonlinear advection term.  Therefore, if the conditions for shock
formation were to arise, our calculation would capture that
shock.\footnote{Since entropy is fixed the compression of our shocks
would be the same as for continuous compression waves. This gives a higher
jump in density for the same jump in pressure than in a proper
dissipative shock. Fortunately, we do not need to contend with this
issue in practice as shocks do not form in our simulations.}

\subsection{Field-line constants}

The poloidal magnetic field is fully described by the azimuthal
component of the vector potential,
\begin{equation}
  B^i = \frac{1}{\sqrt{\gamma}} \epsilon^{ij\phi} \pder{A_\phi}{x^j}\,
  .
\end{equation}

For axisymmetric solutions $A_\phi=\Psi/2\pi$, where $\Psi(x^i)$, the
so-called magnetic flux function, is the total magnetic flux enclosed
by the circle $x^i=$const ($x^i$ being the coordinates of the
meridional plane). Stationary and axisymmetric ideal MHD flows have 5
quantities that propagate unchanged along the magnetic field lines and
thus are functions of $\Psi$ alone. These are $\massint$, the
rest-mass energy flux per unit magnetic flux; $\Omega$, the angular
velocity of magnetic field lines; $l$, the the total angular momentum
flux per unit rest-mass energy flux; $\enint$, the total energy flux
per unit rest-mass energy flux; and $Q$, the entropy per particle.
For cold flows ($Q=0$, $w=\rho c^2$) we have
\begin{equation}
  \massint = \frac{\rho u_p}{B_p}\, ,
\label{chi-def}
\end{equation}  
\begin{equation}
  \Omega=\frac{v^{\hat{\phi}}}{r}-\frac{v_p}{r}
  \frac{B^{\hat{\phi}}}{B_p} \, ,
\label{omega-def}
\end{equation}
\begin{equation}
    l = -\frac{I}{2\pi\massint c}+r u^{\hat\phi}\, ,
\label{angm-def}
\end{equation}
and
\begin{equation}
    \enint = \Gamma \left(1+\sigma\right),
\label{kap-def}
\end{equation}  
where $u_p=\Gamma v_p$ is the magnitude of the poloidal component of
the 4-velocity, $B_p$ is the magnitude of the poloidal component of
the magnetic field, $r$ is the cylindrical radius,
\begin{equation}
   I = \frac{c}{2} r B^{\hat\phi}
\label{I}
\end{equation}  
is the total electric current flowing through a loop of radius $r$,
$\sigma$ is the ratio of the Poynting flux to the matter (kinetic plus
rest-mass) energy flux, and
\begin{equation}
 \Gamma \sigma=-\frac{\Omega I}{2\pi\massint c^3}
\label{sigma-def}
\end{equation}  
is the Poynting flux per unit rest-mass energy flux. (Here and in the
rest of the paper we use a hat symbol over vector indices to indicate
their components in a normalized coordinate basis.)  From
equation~(\ref{kap-def}) it follows that the Lorentz factor $\Gamma$
cannot exceed $\enint$.

\section{Numerical Simulations}
\label{simulations}

To maintain a firm control over the jet's confinement and to prevent
complications related to numerical diffusion of the dense
nonrelativistic plasma from the jet's surroundings, we study outflows
that propagate inside a solid funnel of a prescribed
shape.\footnote{In real astrophysical systems, the shape of the
boundary is determined by the spatial distribution of the pressure or
the density of the confining ambient medium
\citep[e.g.][]{BR74,K82,K94}. The effective ambient pressure
distributions implied by the adopted funnel shapes are considered in
Section~\ref{application}.}  Specifically, we consider axisymmetric
paraboloidal funnels
\begin{eqnarray}
\nonumber z \propto r^a\, ,
\end{eqnarray}
where $z$ and $r$ are the cylindrical coordinates of the funnel wall.
This suggests the utilization of a system of coordinates in which the
funnel wall is a coordinate surface. For a conical jet ($a=1$) we use
spherical coordinates, whereas for jets with $a>1$ we employ
elliptical coordinates $\{\xi,\eta,\phi\}$, where
\begin{equation}
 \xi=rz^{-1/a}
\label{xi}
\end{equation}
and
\begin{equation}
 \eta^2=\frac{r^2}{a} + z^2
\label{eta}
\end{equation}
(see Appendix~\ref{appendix_1} for details).\footnote{The equations are
dimensionalized in the following manner. The unit of length, $L$, is
such that $\eta_i=1\, L$, where the subscript $i$ refers to the inlet
boundary. The unit of time is $T=L/c$. The unit of mass is $M=L^3
B_0^2/4\pi c^2$, where $B_0$ is the dimensional magnitude of the $\eta$
component of magnetic field at the inlet (so the dimensionless magnitude
of $B^{\hat\eta}$ at the inlet is $\sqrt{4\pi}$).  In applications, $L$
is the length scale of the launch region (e.g. the radius of the event
horizon if the jet originates in a black hole), $T$ is the light
crossing time of that region and $B_0$ is the typical strength of the
poloidal magnetic field at the origin.}

We use a Godunov-type numerical code based on the scheme described in
\citet{K99}.  To reduce numerical diffusion we applied parabolic
reconstruction instead of the linear one of the original code. Our
procedure,
in brief, was to calculate minmod-averaged first and second derivatives and use
the first three terms of the Taylor expansion for spatial
reconstruction.
This simple procedure has resulted in a noticeable
improvement in the solution accuracy even though the new scheme is still
not 3rd-order accurate because of the non-uniformity of the grid.

The grid is uniform in the $\xi$ direction (the polar angle direction
when we use spherical coordinates), where in most runs it has a total
of 60 cells.  To check the convergence, some runs were repeated with a
doubled resolution. The cells are elongated in the $\eta$ direction
(the radial direction when we use spherical coordinates), reflecting
the elongation of the funnel. For very elongated cells we observed
numerical instability, so we imposed an upper limit of 40 on the
length/width ratio.

To speed up the simulations, we implemented a sectioning of the
computational grid as described in \citet{KL04}. In each section,
which is shaped as a ring, the numerical solution is advanced using a
time step based on the local Courant condition. It is twice as large
as the time step of the adjacent inner ring and twice as small as the
time step of the adjacent outer ring.  This approach is particularly
effective for conical flows but less so for highly collimated, almost
cylindrical configurations.

\begin{figure*}
\includegraphics[width=85mm]{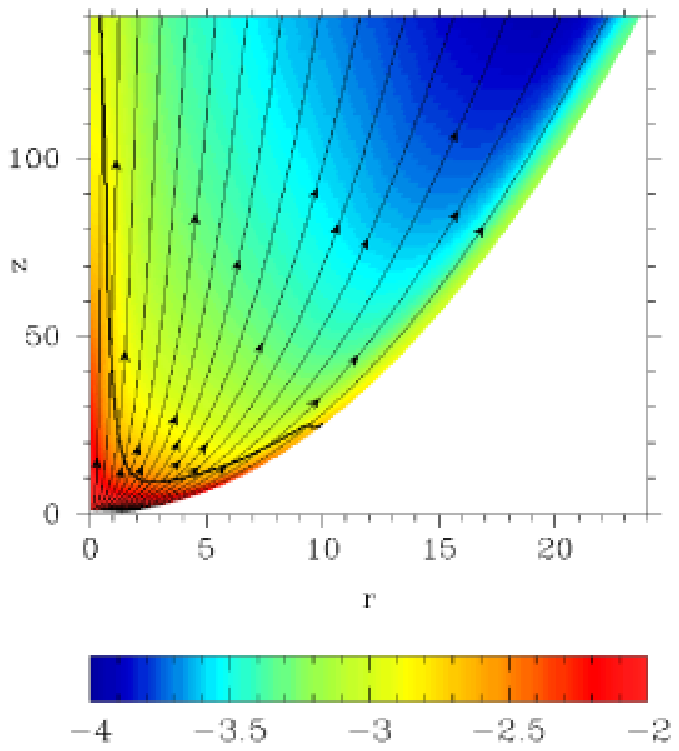}
\includegraphics[width=85mm]{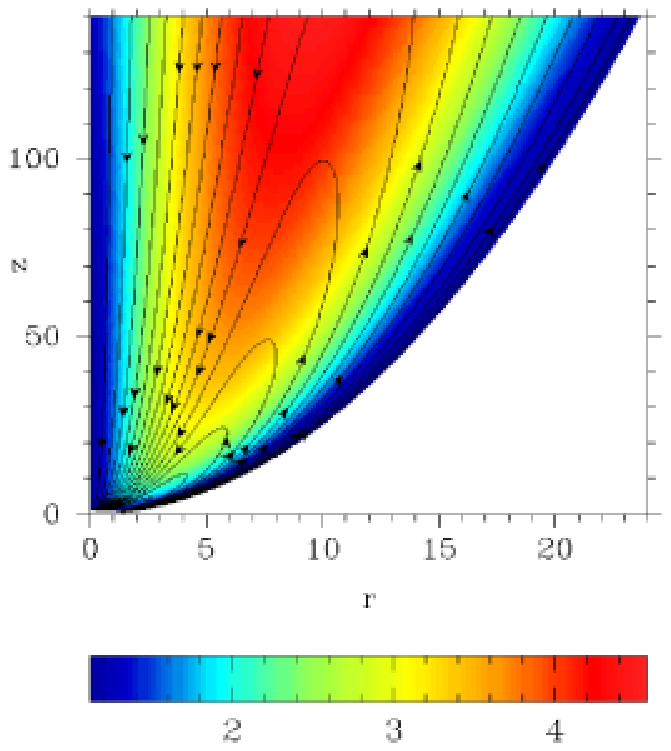}
\includegraphics[width=85mm]{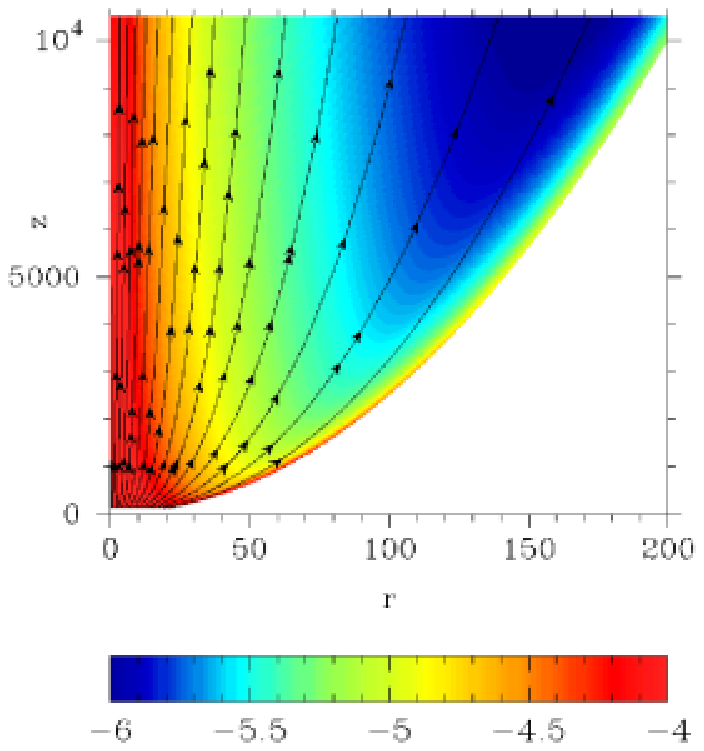}
\includegraphics[width=85mm]{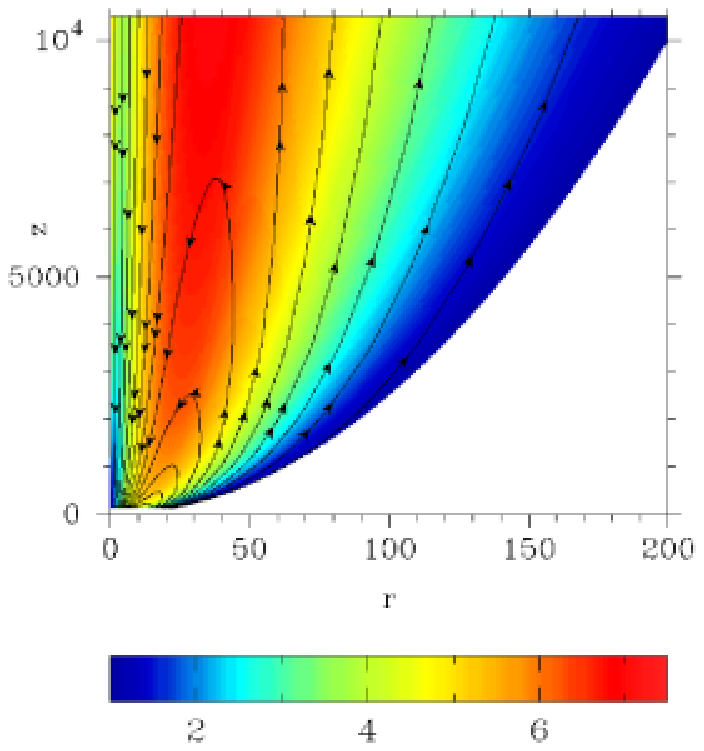}

\caption{Model C1. Left panels show $\log_{10}\Gamma\rho$
(colour), where $\Gamma\rho$ is the jet density as measured in
the laboratory frame, and magnetic field lines. Right panels
show the Lorentz factor (colour) and the current lines. The
thick solid line in the top-left panel denotes the surface where
the flow becomes superfast in the $\eta$ direction. 
The top panels show the solution for the first grid sector,
whereas the bottom panels show the combined solution for the
second and third grid sectors.
}
\label{c1-2d}
\end{figure*}

\begin{figure*}
\includegraphics[width=85mm]{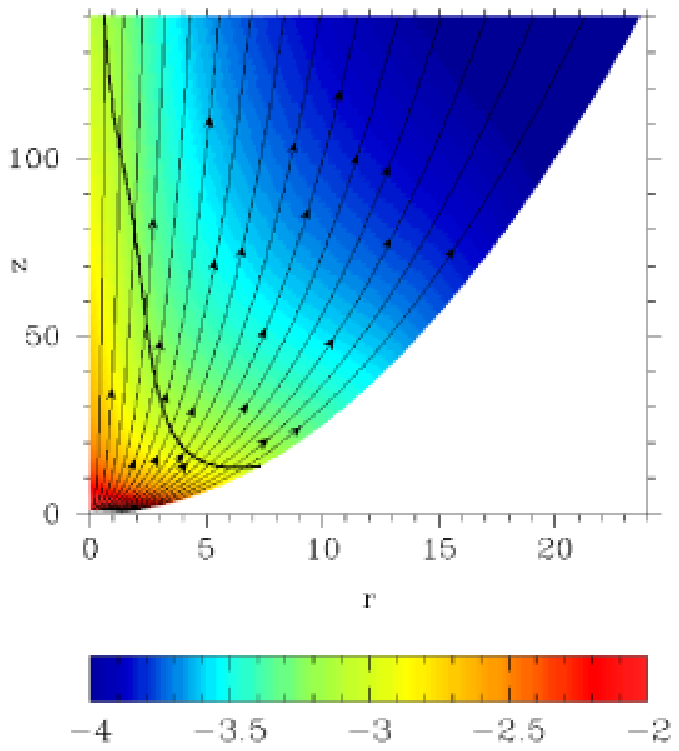}
\includegraphics[width=85mm]{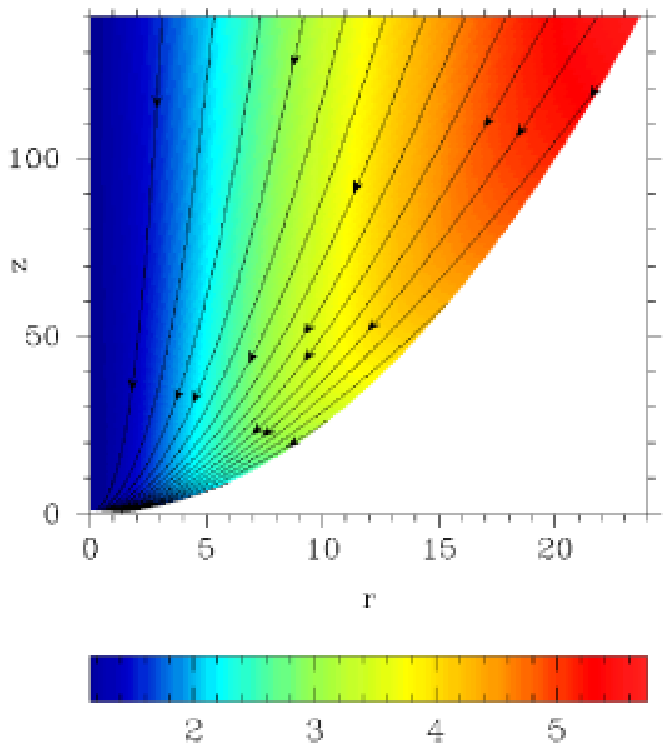}
\includegraphics[width=85mm]{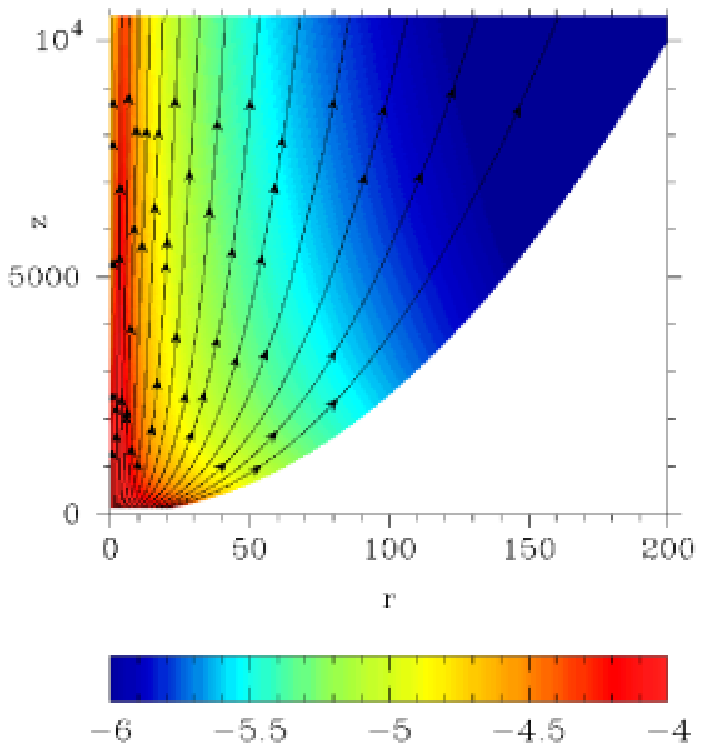}
\includegraphics[width=85mm]{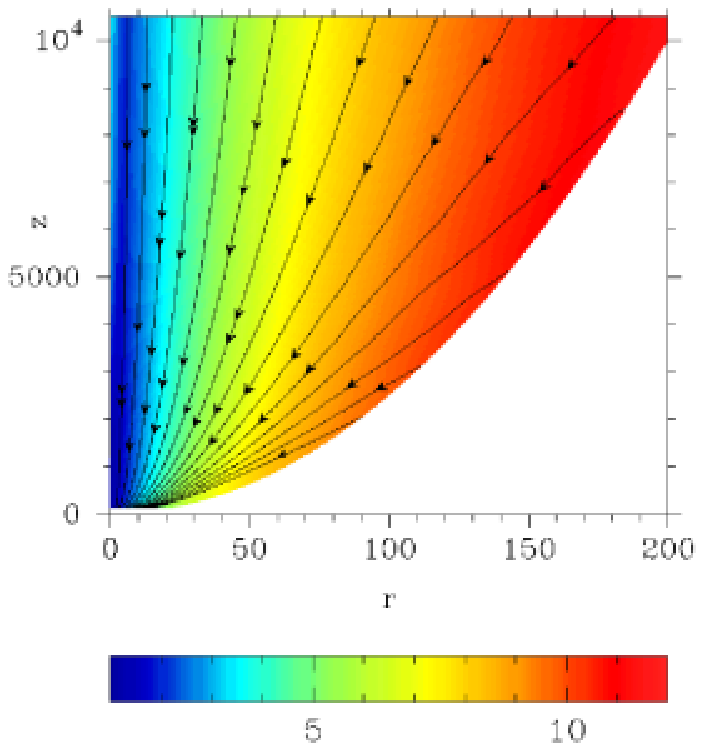}
\caption{Same as Fig.~\ref{c1-2d}, but for Model C2.
}
\label{c2-2d}
\end{figure*}

\begin{figure*}
\includegraphics[width=85mm]{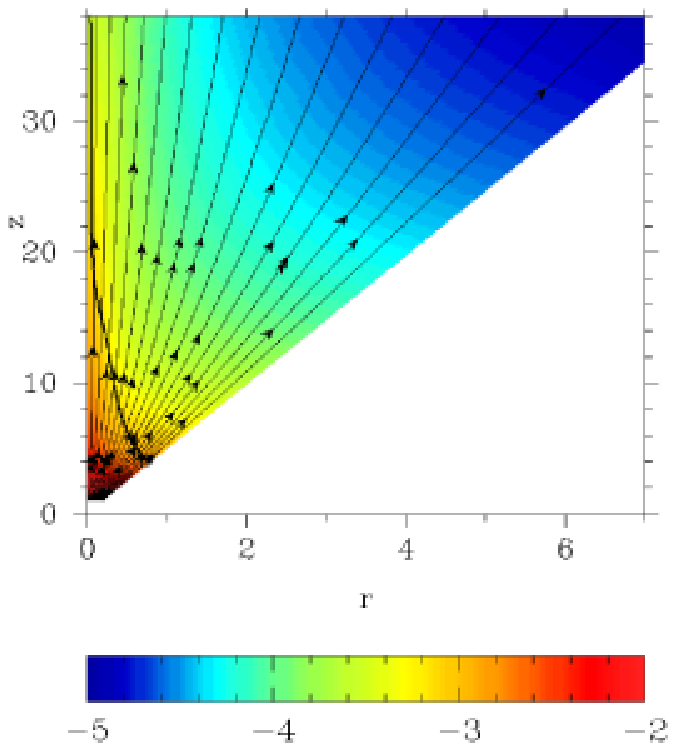}
\includegraphics[width=85mm]{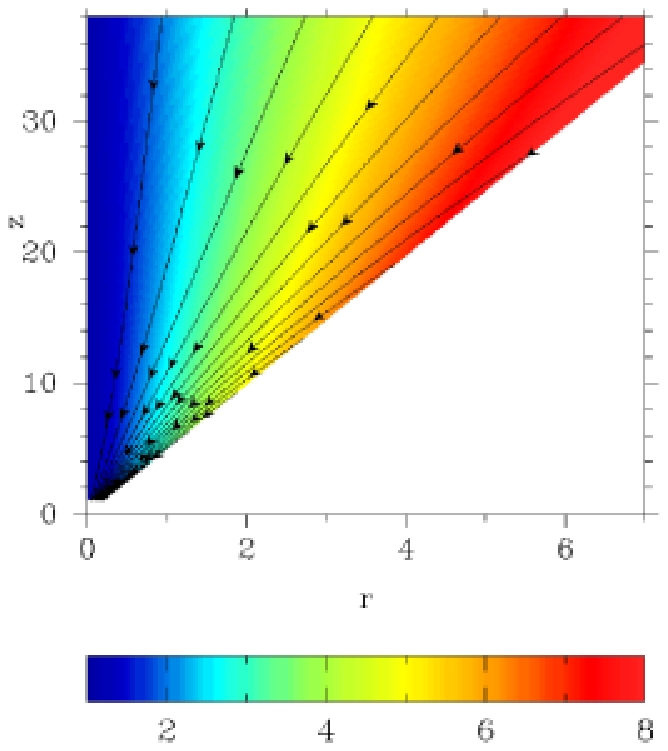}
\includegraphics[width=85mm]{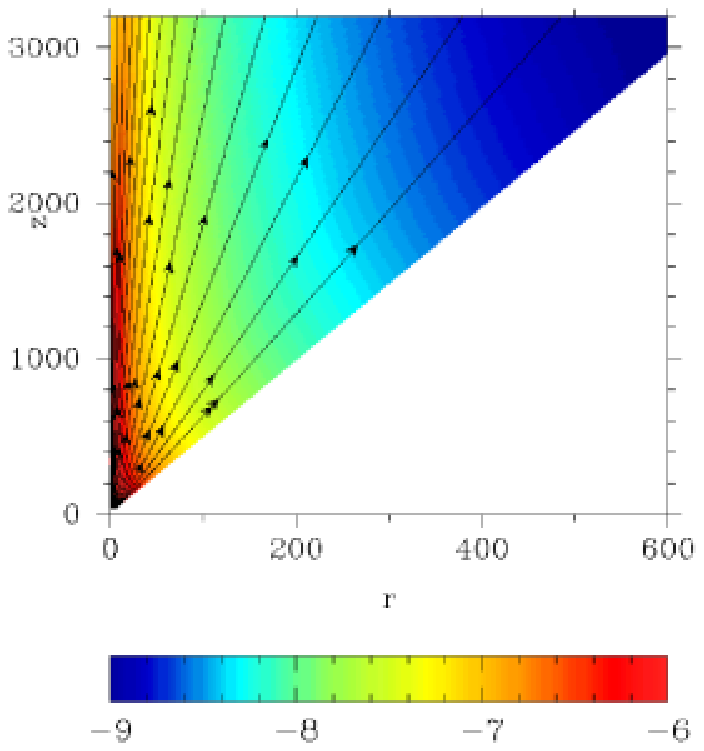}
\includegraphics[width=85mm]{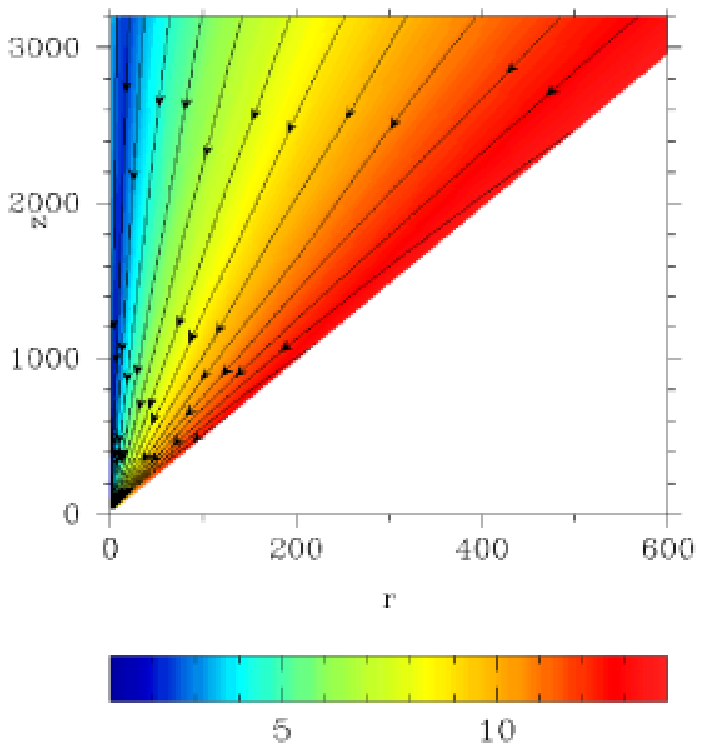}
\caption{Same as Fig.~\ref{c1-2d}, but for Model A2. 
}
\label{a2-2d}
\end{figure*}
                                                            
\subsection{Boundary conditions}

\subsubsection{Inlet boundary}
\label{inlet_section}

We treat the inlet boundary, $\eta_i=1$, as a surface of a perfectly
conducting rotator and consider two rotation laws,
\begin{equation}
\Omega= \Omega_0
\label{Omega0}
\end{equation}
and
\begin{equation}
\Omega= \Omega_0 [1-3(\xi/\xi_j)^2+2(\xi/\xi_j)^3]\, ,
\label{Omega}
\end{equation}
where $\xi_j$ marks the jet boundary.  The angular velocity profile is
directly related to the distribution of the return electric current in
the jet (see equation~\ref{I-Bp} below). In fact, the current is driven
by the electric field associated with the rotating poloidal field, and
charge conservation requires the circuit to eventually close.  In
the case of constant $\Omega$ the return current flows over the jet
boundary.  For the rotation law (\ref{Omega}) it is distributed over the
jet body as a volume current, the current changing sign at
$\xi\simeq\xi_j/2$.  Thus, we cover the two generic types of electric
current distribution.  
The solid-body rotation law provides a very good
description of the behaviour of magnetic field lines that thread the horizon
of a black hole. This choice is therefore entirely appropriate for the
black-hole theory of relativistic AGN jets. The differential rotation law is
more suitable to the accretion-disk theory,  although it admittedly does
not correspond to a realistic velocity field (which is hard to model given
the limitations of our numerical technique).

The condition of perfect conductivity allows us to fix the azimuthal
component of the electric field and the $\eta$ component of the
magnetic field:
\begin{equation}
  E_\phi=0, \quad B^{\hat\eta}=B_0 \text{at} \eta=\eta_i\, .
\end{equation}
From the first of these conditions we derive
\begin{equation}
v^{\hat\xi} = \frac{v^{\hat\eta}}{B^{\hat\eta}} B^{\hat\xi}
\label{v_xi_b}
\end{equation}
and (using equation \ref{omega-def})
\begin{equation}
v^{\hat{\phi}} = r \Omega + \frac{v^{\hat\eta}}{B^{\hat\eta}}
B^{\hat{\phi}}\, .
\label{v_phi_b}
\end{equation}
We have also experimented with non-uniform distributions of the magnetic
field, in particular with $B^{\hat\eta}$ decreasing with $\xi$. The
results were not significantly different as the
field distribution downstream of the inlet underwent a rapid
rearrangement that restored the transverse force balance.

On the assumption of a cold (i.e. zero thermal energy) jet, the flow
at the inlet boundary is necessarily super--slow-magnetosonic. This
means that both the density and the radial component of the velocity
can be prescribed some fixed values:
\begin{eqnarray}
\nonumber \rho=\rho_0\, , \quad v^{\hat\eta}=v_{p_{0}}\, .
\end{eqnarray}
In the simulations we used $v_{p_{0}}=0.5\, c$, which was a choice of 
convenience. On the one hand, this value is sufficiently small  
to insure that the flow at $\eta=1$ is sub-Alfv\'enic and hence that
the Alfv\'en and fast-magnetosonic critical surfaces are located
downstream of the inlet boundary. On the other hand, it is large enough 
to promote a rapid settlement to a steady state (keeping in mind that the
speed of a steady-state flow remains constant along the symmetry axis).
Because of the sub-Alfvenic nature of the inlet flow, we cannot fix the
other components of the magnetic field and the velocity -- they are to
be found as part of the global solution.  Following the standard
approach we extrapolate $B^{\hat\phi}$ and $B^{\hat\xi}$ from the
domain into the inlet boundary cells.  We then compute $v^{\hat\phi}$
and $v^{\hat\xi}$ from equations~(\ref{v_xi_b}) and~(\ref{v_phi_b}).

In the case of differential rotation the magnitude of the angular
velocity is chosen in such a way that the Alfv\'en surface of the jet
is near the jet origin, its closest point being located at a distance
of $\sim 1.5$ times the initial jet radius from the inlet surface.  In
the case of solid-body rotation the Alfv\'en surface almost coincides
with the light cylinder, whose radius $r_{lc} \equiv c/\Omega$ is only
$50\%$ larger than the initial jet radius.

The inlet density is chosen so that all jets have very similar values
of $\enint$ and $\sigma$. In particular, for the models with uniform
$\Omega$ we have $\enint_{\rm max}\simeq 18$, and for the models with
non-uniform $\Omega$ we have $\enint_{\rm max}\simeq 12$.

\subsubsection{Other boundaries}

The computational domain is always chosen to be long enough for the
jet to be super--fast-magnetosonic when it approaches the outlet
boundary $\eta=\eta_o$. This justifies the use of radiative boundary
conditions at this boundary (i.e. we determine the state variables of
the boundary cells via extrapolation of the domain solution).

At the polar axis, $\xi=0$, we impose symmetry boundary conditions for
the dependent variables that are expected to pass through zero there,
\begin{eqnarray}
\nonumber f(-\xi)=-f(\xi)\, .
\end{eqnarray}
These variables include $B^{\hat\xi}$, $B^{\hat\phi}$, $u^{\hat\xi}$
and $u^{\hat\phi}$.  For other variables we impose a ``zero second
derivative'' condition,
\begin{eqnarray}
\nonumber \partial^2{f}/\partial{\xi^2} = 0\, ,
\end{eqnarray}
which means that we use linear interpolation to calculate the values of these
variables in the boundary cells.

We do this in order to improve the numerical representation of a
narrow core that develops in all cases as a result of the magnetic
hoop stress. Within this core the gradients in the $\xi$ direction are
very large and the usual zero-gradient condition, $f(-\xi)=f(\xi)$,
results in increased numerical diffusion in this region. We have
checked that this has a noticeable effect only on the axial region and
that the global solution does not depend on which of these two
conditions is used.

At the wall boundary, $\xi=\xi_j$, we use a reflection condition,
\begin{eqnarray}
\nonumber f(\xi_j+\Delta \xi)=-f(\xi_j-\Delta\xi) \, ,
\end{eqnarray}
for $B^{\hat\xi}$ and $u^{\hat\xi}$ and a zero-gradient condition for
all other variables.

\subsection{Initial setup}
\label{setup}

The initial configuration corresponds to a non-rotating, purely poloidal
magnetic field with approximately constant magnetic pressure across the
funnel. The plasma density within the funnel is set to a small value so
that the outflow generated at the inlet boundary can easily sweep it
away. In order to speed this process up the $\eta$ component of velocity
inside the funnel is set equal to $0.7\, c$, whereas the $\xi$ component
is set equal to zero.

\begin{figure*}
\includegraphics[width=55mm,angle=-90]{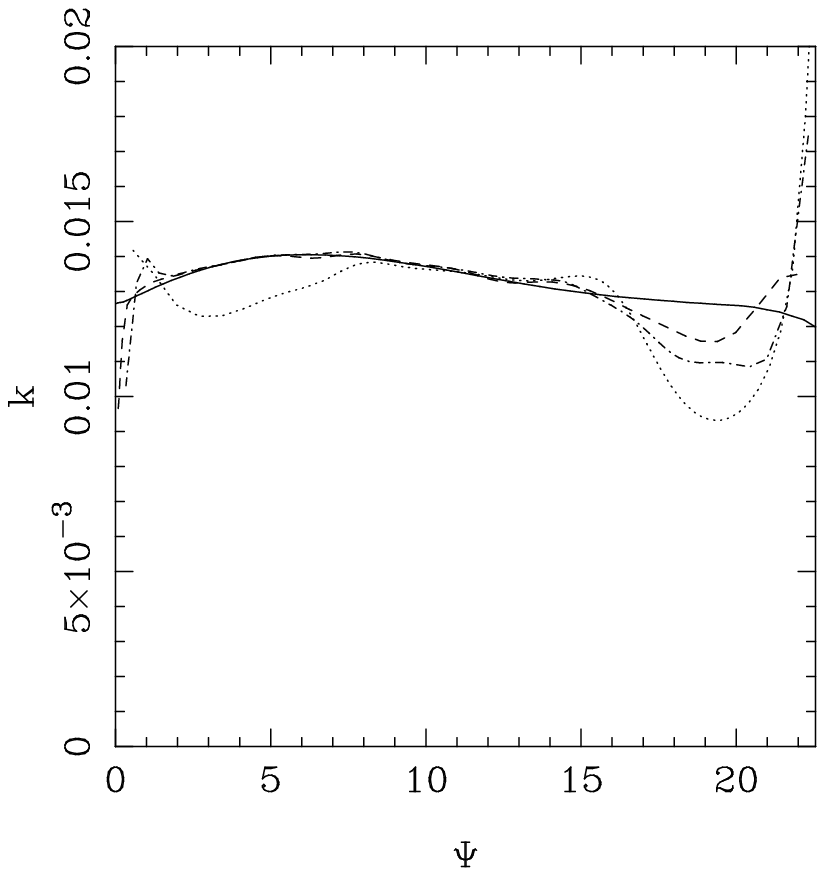}
\includegraphics[width=55mm,angle=-90]{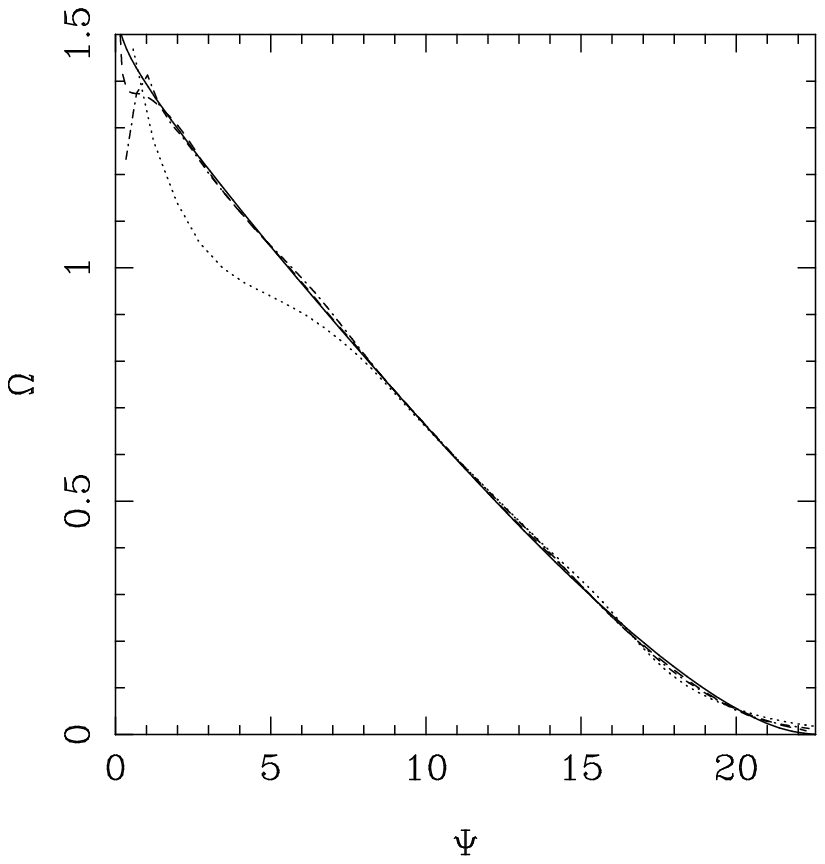}
\includegraphics[width=55mm,angle=-90]{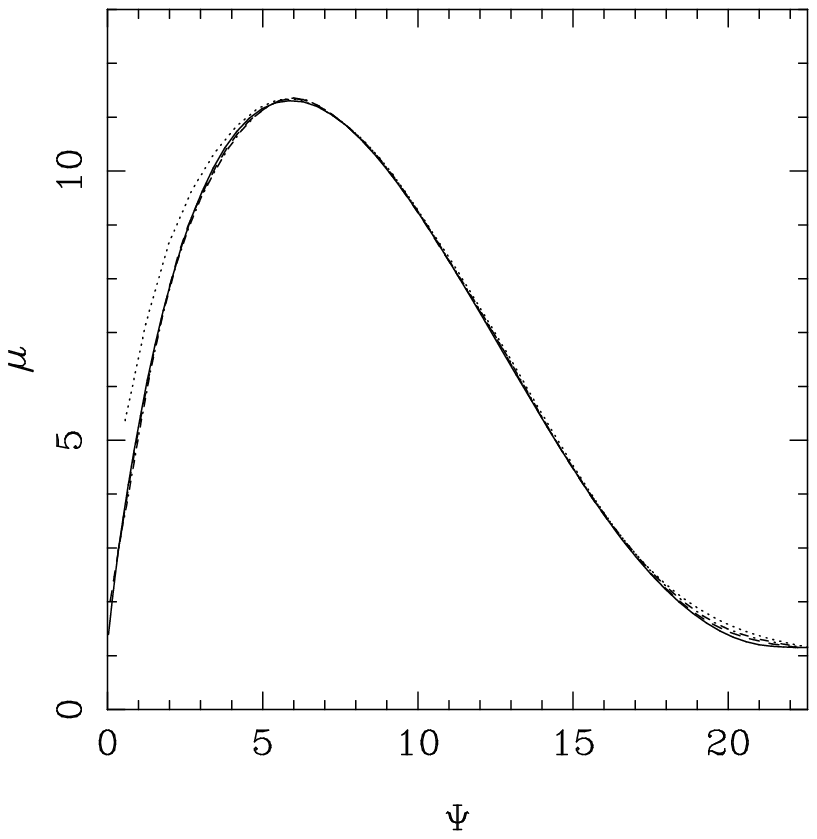}
\includegraphics[width=55mm,angle=-90]{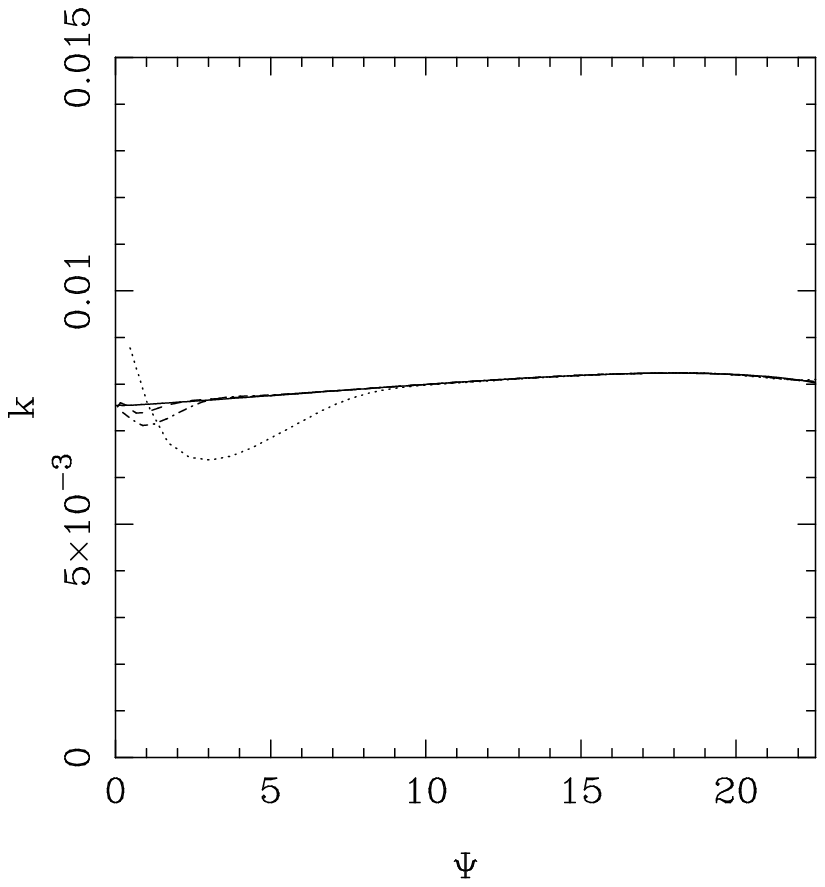}
\includegraphics[width=55mm,angle=-90]{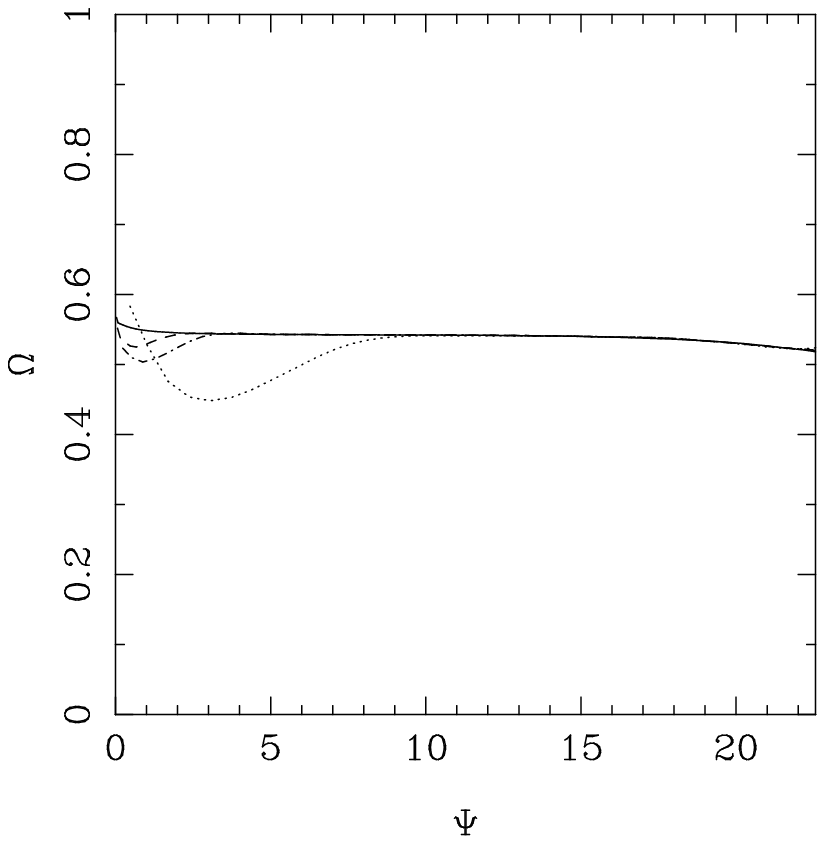}
\includegraphics[width=55mm,angle=-90]{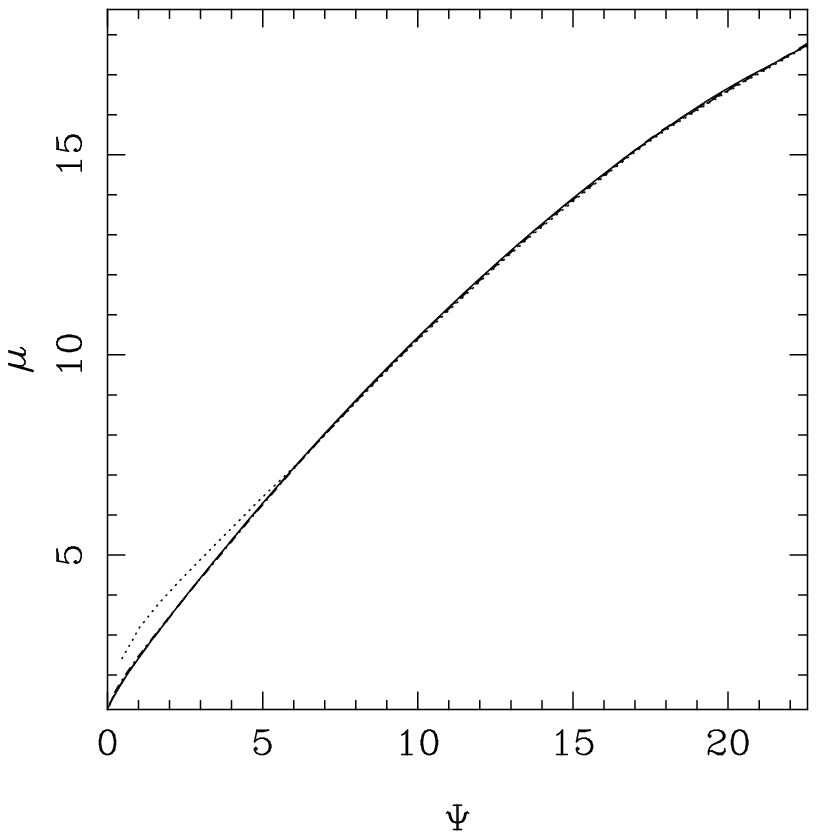}
\includegraphics[width=55mm,angle=-90]{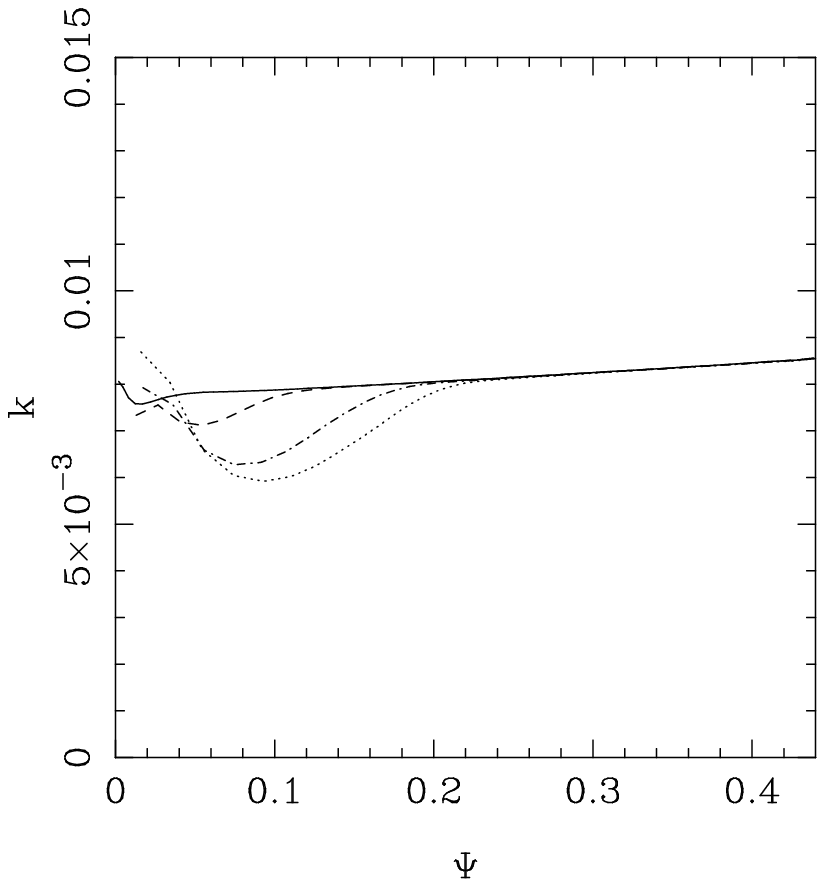}
\includegraphics[width=55mm,angle=-90]{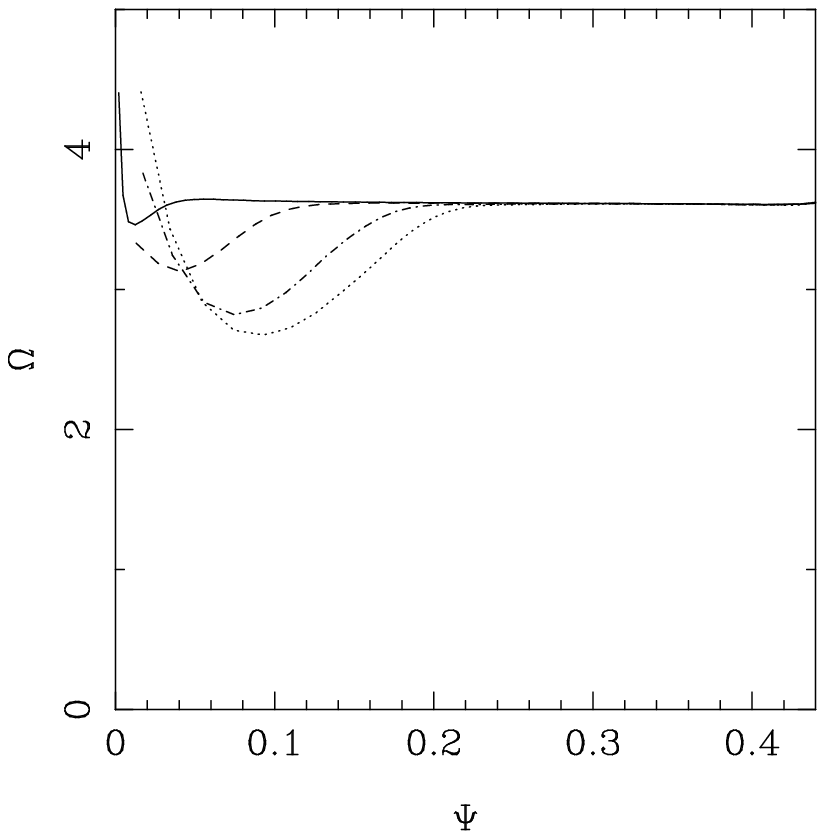}
\includegraphics[width=55mm,angle=-90]{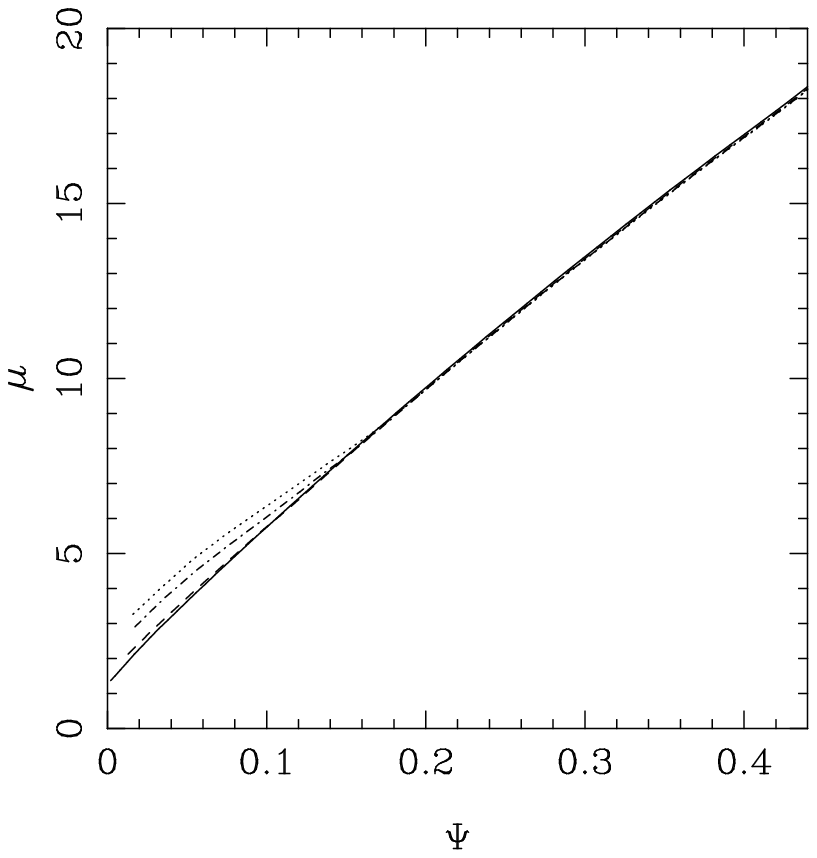}

\caption{Flow constants $\massint(\Psi)$, $\Omega(\Psi)$ and
$\enint(\Psi)$ at different distances from the source for models C1
(top row), C2 (middle row) and A2 (bottom row).  In all cases the
solid lines show the constants at the injection surface ($\eta=1$).
The deviations from these values further downstream are due to a
gradual accumulation of numerical errors. For models C1 and C2 the
dashed line corresponds to $\eta=10^2$, the dash-dotted line to
$\eta=10^3$, and the dotted line to $\eta=10^4$.  For model A2 the
dashed line corresponds to $\eta=2\times10^2$, the dash-dotted line to
$\eta=2\times10^3$, and the dotted line to $\eta=2\times10^4$.  
}
\label{constants}
\end{figure*}

\begin{figure}
\includegraphics[width=77mm,angle=-90]{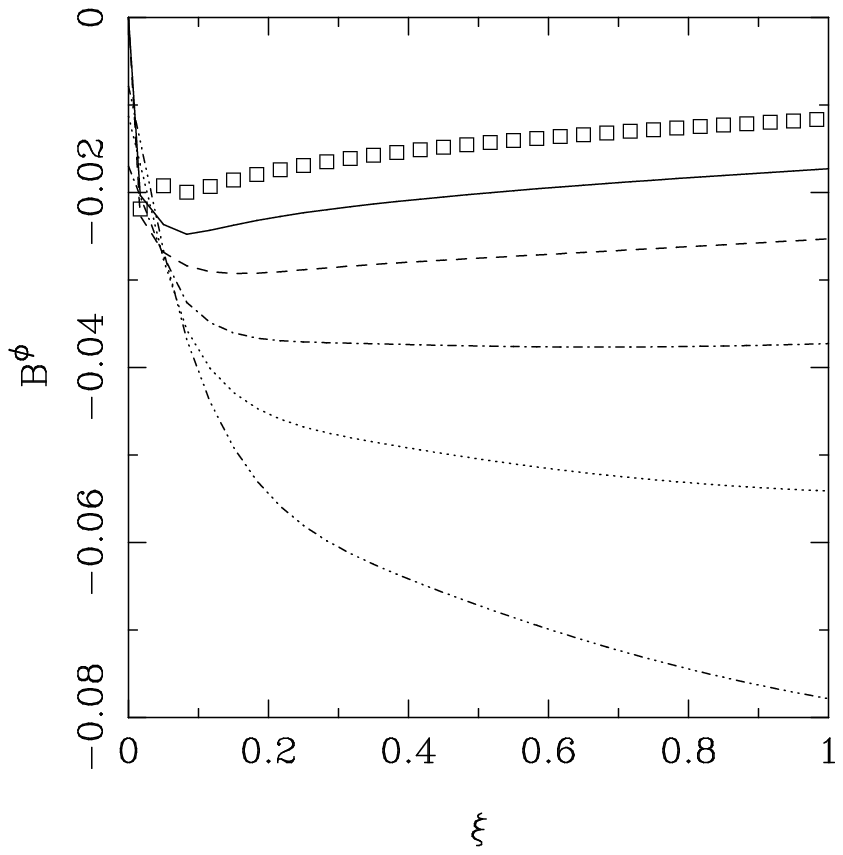}
\caption{ Development of the line current in model C2. The figure
shows the azimuthal magnetic field at $\eta=200,\, 400,\, 800,\,
1600,\, 3200$ and 6400 (from bottom to top). The solution at
$\eta=6400$ is plotted as squares.  }
\label{cu-core}
\end{figure}
\subsection{Grid extensions}
 
The inner rings of the grid, where the grid cells are small and so is
the time step, are the computationally most intensive regions of the
simulation domain. If we kept computing these inner rings during the
whole run then we would not be able to advance very far from the jet
origin.  Fortunately, the trans-sonic nature of the jet flow allows us
to cease computations in the inner region once the solution there
settles to a steady state.  To be more precise, we cut the funnel
along the $\xi$-coordinate surfaces into overlapping sectors with the
intention of computing only within one sector at any given time,
starting with the sector closest to the inlet boundary.  Once the
solution in the ``active'' sector settles to a steady state we switch
to the subsequent sector, located further away from the inlet. During
the switch the solution in the outermost cells of the active sector is
copied into the corresponding inner boundary cells of the subsequent
sector.  During the computation within the latter sector these inner
boundary cells are not updated.  Surely, this procedure is justified
only when the flow in a given sector cannot communicate with the flow
in the preceding sector through hyperbolic waves, and thus we need to
ensure that the Mach cone of the fast-magnetosonic waves points
outward at the sector interfaces. This condition can be written as
\begin{eqnarray}
\label{ext_criterion}
\left(\Gamma {v^{\hat\eta}}\right)^4- \left(\Gamma
{v^{\hat\eta}}\right)^2 \left( \frac{b^2}{4 \pi w/c^2} +\frac{ c_s^2 }
{1-{c_s^2/c^2} }\right)+ \nonumber \\ + \frac{ c_s^2 } {1-{c_s^2/c^2}
} \frac{\left(B^{\hat\eta}\right)^2 }{4 \pi w/c^2}
\left(1-\frac{r^2}{r_{lc}^2}\right) > 0 \,,
\end{eqnarray}
where $c_s^2=spc^2/w$ (see equation~\ref{eos}) and $b=(B^2-E^2)^{1/2}$
is the magnetic field magnitude in the fluid frame (see
Appendix~\ref{appendix_2} for details).  In the cold limit this
reduces to
\begin{eqnarray}
\label{criterion_cold}
v^{\hat\eta}> \left
[{1-\left({v^{\hat\xi}}/{c}\right)^2-\left(v^{\hat\phi}/c\right)^2}
\right ]^{1/2} c_f \,,
\end{eqnarray}
where $c_f=b/({4\pi\rho +b^2/c^2})^{1/2}$ is the isotropic fast speed
in the fluid frame.  Thus, the jet has to be super-fast in the $\eta$
direction at the sector interfaces.  In Figs.~\ref{c1-2d}--\ref{a2-2d}
the location of the surface where $v^{\hat\eta}=c_f$ is shown by a
thick solid line: to the right of this line $v^{\hat\eta}>c_f$. One
can see that the transition to the super-fast regime occurs well
inside the first sector.  (Note that when $v^{\hat\eta}>c_f$ the
inequality \ref{criterion_cold} is satisfied.)

In these simulations we normally used 4 or 5 sectors, with each
additional sector being ten times longer than the preceding one. This
technique has enabled us to reduce the computational time by up to
three orders of magnitude, depending on the funnel geometry.  Although
the grid extension can in principle be continued indefinitely, there
are other factors that limit how far along the jet one can advance in
practice. Firstly, once the paraboloidal jets become highly collimated
the required number of grid cells along the jet axis increases, and
each successive sector becomes more expensive than the previous one.
Secondly, computational errors due to numerical diffusion gradually
accumulate in the downstream region of the flow and the solution
becomes progressively less accurate (see Fig.~\ref{constants}).

\section{Results}
\label{results}

Models A, B, C and D have geometrical power indices $a=1,\, 3/2,\, 2$
and $3$, respectively. Further classification is based on the rotation
law: models A1--D1 have non-uniform rotation, whereas models A2--D2
have uniform rotation.

Models with different power indices but the same rotation law show
remarkably similar properties. Thus, it is sufficient to show only one
of them in greater detail. For this purpose we selected models C1, C2
and A2.

Fig.~\ref{c1-2d} shows the distribution of the Lorentz factor, the
lab-frame rest-mass density, the poloidal magnetic field, and the
poloidal electric current for model C1. The top panel presents the
solution in the innermost grid sector whereas the bottom panels show
the solution in the second and third sectors ``glued'' together.  The
density distribution as well as the magnetic field lines clearly
indicate that the jet develops a core where the magnetic surfaces
become almost cylindrical.\footnote{The isodensity contours are, in
fact, more strongly collimated than the magnetic field lines (or flow
streamlines), a trait already identified previously in nonrelativistic
\citep{S95} and moderately relativistic \citep{Li96b} MHD disk
outflows.}  The core is produced by the hoop stress of the toroidal
magnetic field that is wound up in the main body of the jet due by the
rotation at its base. The ratio of the core radius to the jet radius
decreases monotonically with increasing distance from the source until
the core eventually becomes unresolved on the grid. The solution then
develops what looks like an axial line current. (Such behaviour is
also observed in models A2--D2; see Fig.~\ref{ahu-b}).

Near the jet boundary, where the angular velocity of the magnetic
field lines in model C1 vanishes, the azimuthal magnetic field
component is weak and the equilibrium is supported in part by the
poloidal field. If the flow were self-similar, one would have $B_p
\propto r_j^{-2}$, $B^{\hat\phi} \propto r_j^{-1}$, and the pressure
of the azimuthal field in the main body of the jet would eventually
become much larger than the pressure of the poloidal field in the
boundary sheath, leading to a loss of force balance.  In reality, the
flow adjusts through a progressive compression of the
sheath. Consequently a thin layer of surface current gradually
develops as the distance from the source increases. This is why some
current lines in the bottom-right panel of Fig.~\ref{c1-2d} appear to
terminate at the jet surface.

\begin{figure*}
\includegraphics[width=77mm,angle=-90]{figures/cu-lines.eps}
\includegraphics[width=77mm,angle=-90]{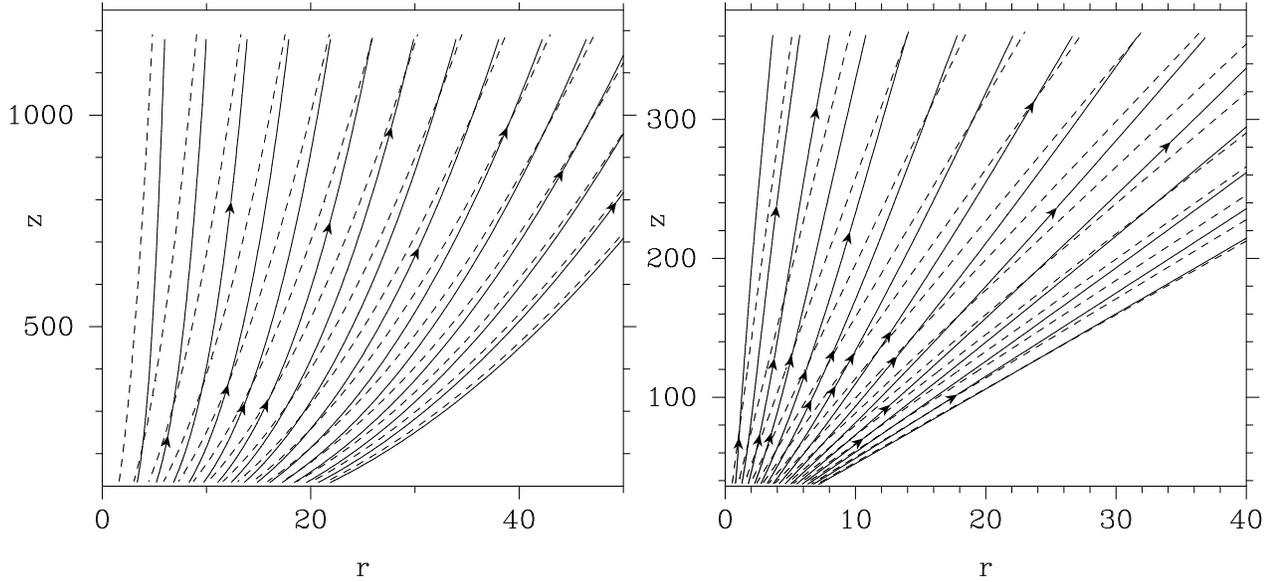}
\caption{Poloidal magnetic field lines (solid) and $\xi=$const
coordinate lines (dashed) for models C2 (left panel) and A2 (right
panel). In all models the magnetic field lines show faster collimation
than the coordinate lines.  }
\label{lines}
\end{figure*}

\begin{figure*}
\includegraphics[width=55mm,angle=-90]{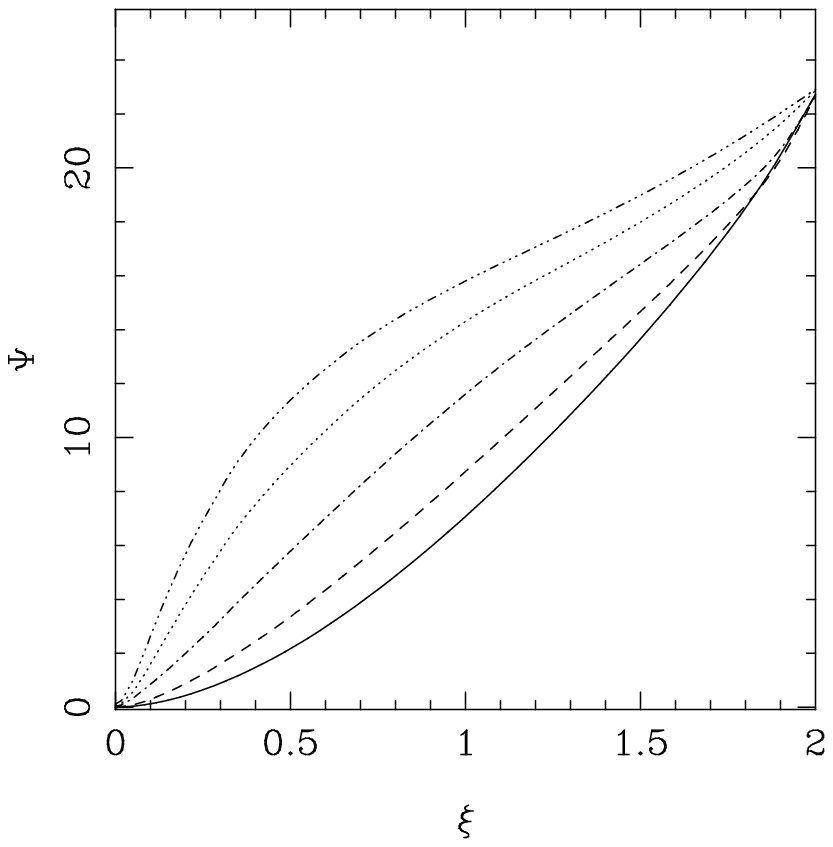}
\includegraphics[width=55mm,angle=-90]{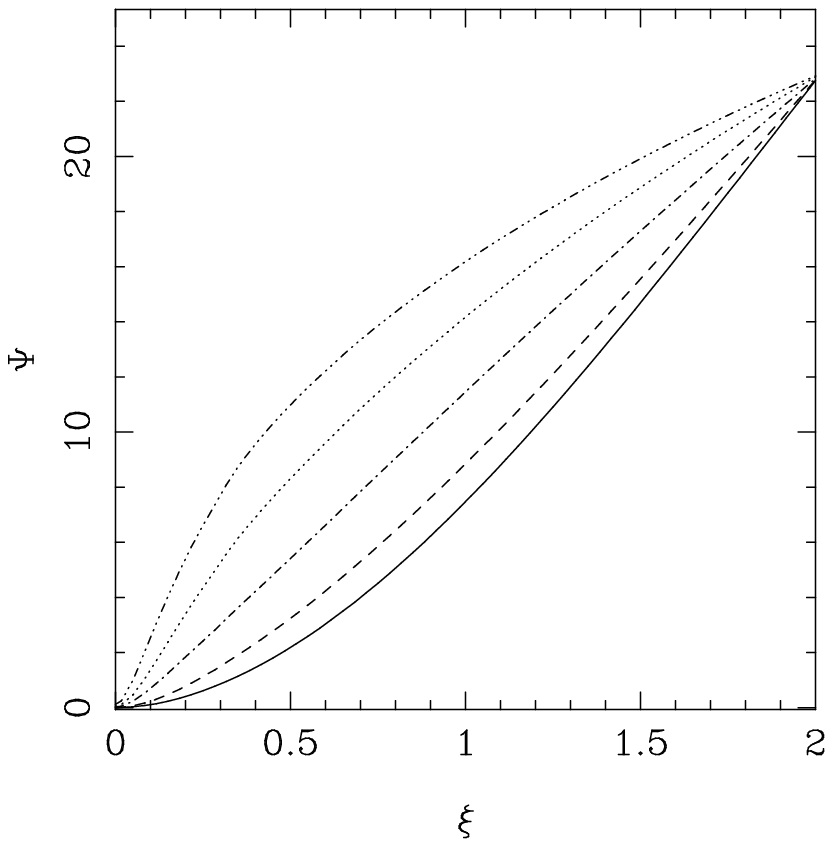}
\includegraphics[width=55mm,angle=-90]{figures/au-mflux.eps}
\caption{Evolution of the magnetic flux distribution across the jet
with distance from the inlet. {\it Left panel}: model C1. {\it Middle
panel}: model C2.  In both cases the solid line corresponds to
$\eta=10$, the dashed line to $\eta=10^2$, the dash-dotted line to
$\eta=10^3$, the dotted line to $\eta=10^4$ and the
dash--triply-dotted line to $\eta=10^5$.  {\it Right panel:} model
A2. The solid line corresponds to $\eta=1$, the dashed line to
$\eta=30$, the dash-dotted line to $\eta=3\times10^2$, the dotted line
to $\eta=3\times10^3$ and the dash--triply-dotted line to
$\eta=3\times10^4$.
Note that in the conical case we use the spherical coordinate
$\theta= \arctan \xi $ (in radians) 
rather than 
the $\xi$ coordinate.
}
\label{mflux}
\end{figure*}
                                                                              
The most effective acceleration of the C1 jet occurs at intermediate
cylindrical radii, where the angular velocity of the field lines reaches
a maximum and the poloidal electric current flows across the jet (so the
$(1/c) \vpr{j_p}{B_\phi}$ force is maximized). Note that this aspect of
the jet behaviour could not have been studied with the help of
self-similar models, in which the current lines do not change direction
within the jet.  At comparatively small distances from the inlet
boundary the maximum acceleration occurs at $r_{\rm {max}}\simeq0.5\, r_j$,
but further downstream $r_{\rm {max}}\simeq 0.25\, r_j$.  This is explained
by a more effective collimation in the inner region of the jet than at
the jet boundary (see discussion following equation~\ref{S} in
Section~\ref{theory}).

A careful inspection of the velocity field in the lower-right panel of
Fig.~\ref{c1-2d} reveals an additional region of effective acceleration
near the jet axis for $z \geq 10^3$.  This acceleration, however, is
unphysical as it is caused by numerical diffusion/dissipation in the
core that result from large gradients of the flow variables that develop
there. The gradual growth of errors in this region is clearly seen in
Fig.~\ref{constants}, which shows the flow constants as functions of
$\Psi$ at various distances from the source. Beyond $z=10^4$ the errors
become unacceptably large and this makes further continuation of the
solution via grid extension meaningless.  We note in this connection
that, even in the absence of exact analytic solutions, the existence of
flow constants makes the jet problem a very useful one for testing RMHD
codes and assessing their performance.

\begin{figure*}
\includegraphics[width=55mm,angle=-90]{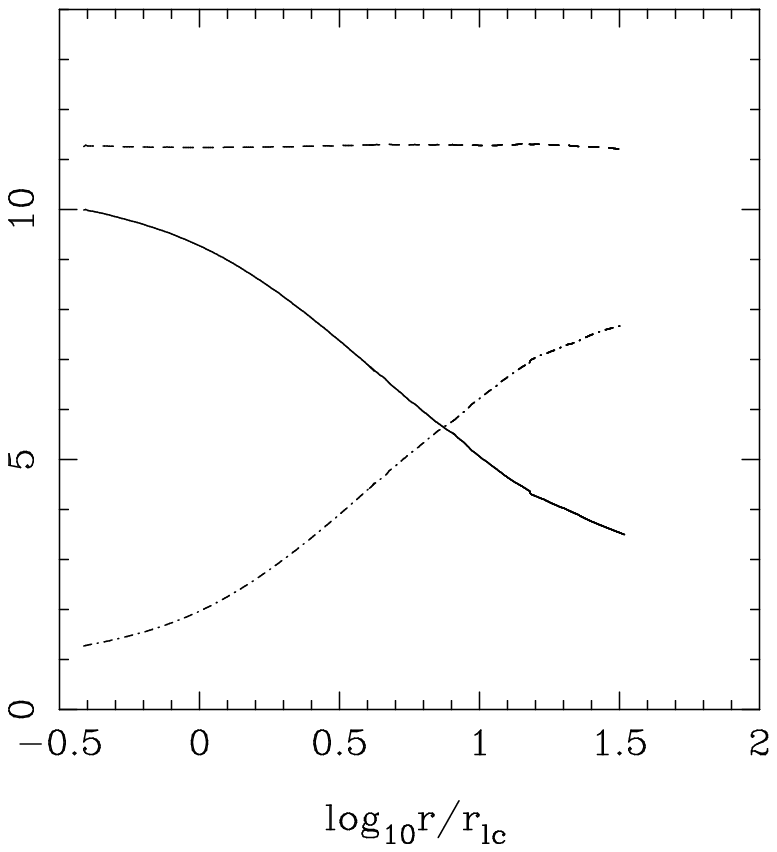}
\includegraphics[width=55mm,angle=-90]{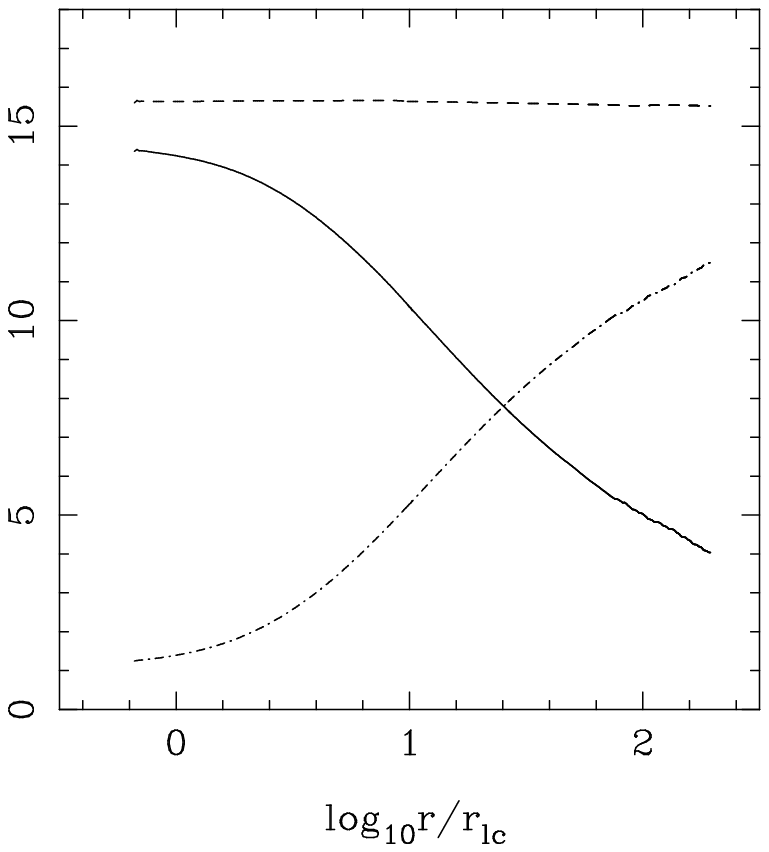}
\includegraphics[width=55mm,angle=-90]{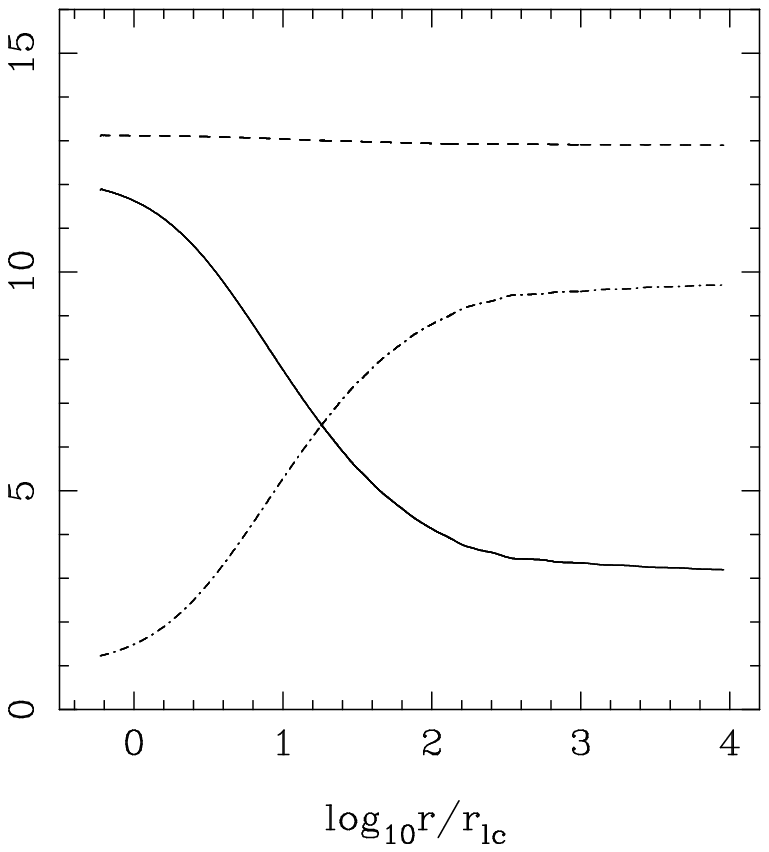}
\caption{$\Gamma\sigma$  (solid line), $\enint$  (dashed line) and  
$\Gamma$ (dash-dotted line) along a magnetic field line 
as a function of cylindrical radius for models C1 (left panel), C2
(middle panel) and A2 (right panel).
}
\label{sigma}
\end{figure*}

Fig.~\ref{c2-2d} shows the distribution of the Lorentz factor,
the lab-frame rest-mass density, the poloidal magnetic field and the
poloidal electric current for model C2. One can see that a core still
develops but that a boundary sheath is no longer present. This is
because the uniform rotation of the magnetic field lines in this model
ensures an effective generation of azimuthal magnetic field all the
way up to the jet boundary. Fig.~\ref{cu-core} shows the development
of an axial line current in this solution, a result of the gradual
decrease of the core radius relative to the jet radius (similar to
what is seen in model C1).  Note, however, that the light cylinder is
unresolved at the distance where the line current is observed. Thus,
what looks like a line current could be a smoothly distributed current
inside the light cylinder.

Inside the jet the electric current flows inward everywhere and
current closure is achieved via a surface current. The radial
component of the current peaks near the boundary, resulting in a
higher $(1/c) \vpr{j_p}{B_\phi}$ force and a more effective plasma
acceleration in this region.

As in the C1 solution, the numerical errors in model C2 grow most
rapidly near the jet axis (see Fig.~\ref{constants}), although they
are somewhat smaller in this case. Moreover, the most interesting
region of the flow, where the acceleration is most effective, is now
far from the axis and does not suffer from these errors as much as in
model C1.  This feature is characteristic not only of models C but of
all the other models as well.  For this reason we decided to focus our
attention on the models with uniform rotation, A2--D2, and in the rest
of this section we present results mainly for these solutions. This
choice is further motivated by the fact that models with a non-uniform
rotation do not seem to exhibit any significant differences with
respect to the uniform-rotation models besides those that we have
already described.

\begin{figure*}
\includegraphics[width=55mm,angle=-90]{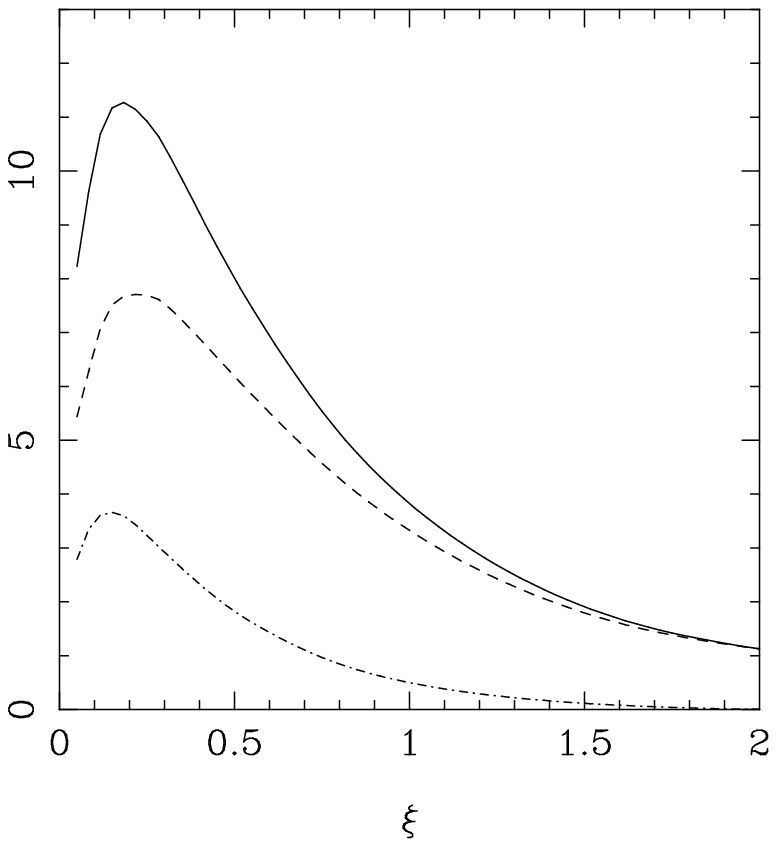}
\includegraphics[width=55mm,angle=-90]{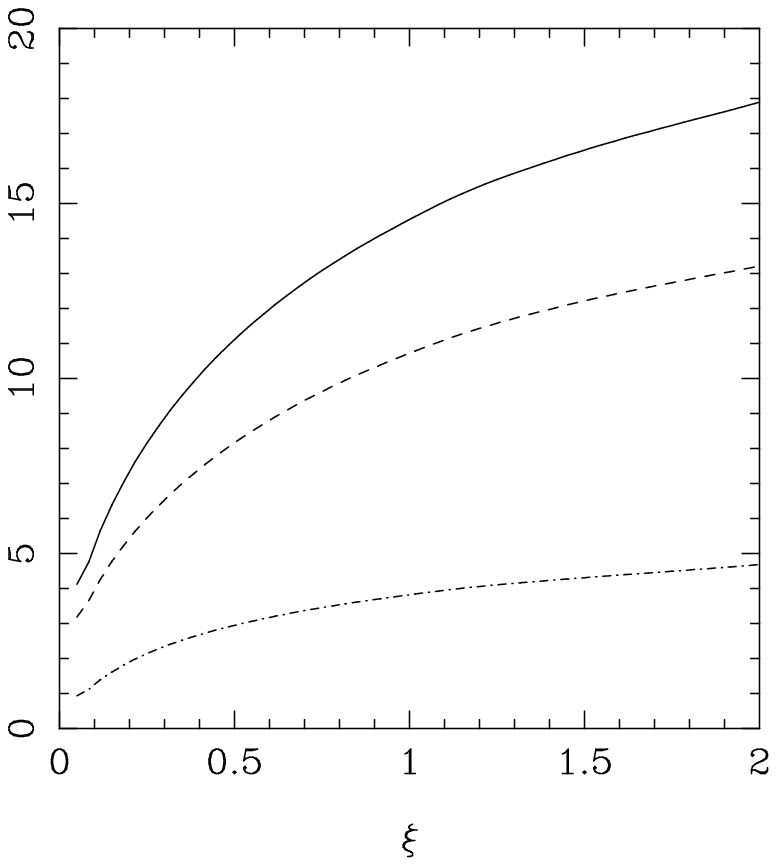}
\includegraphics[width=55mm,angle=-90]{figures/au-cross.eps}
\caption{Distribution of $\enint$ (solid line), $\Gamma$ (dashed line) and 
$\Gamma\sigma$ (dash-dotted line) across the jet for models 
C1 (left panel) and C2 (middle panel) at $\eta=10^5$, and for model A2
(right panel) at $\eta=2\times10^3$. 
}
\label{cross}
\end{figure*}

Given the results of previous analytical and numerical studies, which
suggested poor self-collimation of relativistic magnetized flows (see
references in Section~\ref{theory}), one could have expected the
magnetic flux surfaces to almost mirror the imposed shape of the jet
boundary. However, our results indicate that the outflows collimate
significantly faster, and that this property is manifested not only by
jets with paraboloidal boundaries but also by the ones that are confined
by a conical wall (see Fig.~\ref{a2-2d}).  Fig.~\ref{lines} shows the
magnetic flux surfaces and the coordinate surfaces $\xi = \mbox{const}$
for models A2 and C2. In both cases the magnetic flux surfaces clearly
do not diverge as fast as the coordinate surfaces. This effect is
further demonstrated by Fig.~\ref{mflux}, which shows the evolution of
the magnetic flux distribution across these jets (as well as the jet of
model C1) with distance from the origin. It is seen that the magnetic
flux becomes progressively more concentrated toward the symmetry axis as
the flow moves further downstream.
                                                                             
\begin{figure*}
\includegraphics[width=77mm,angle=-90]{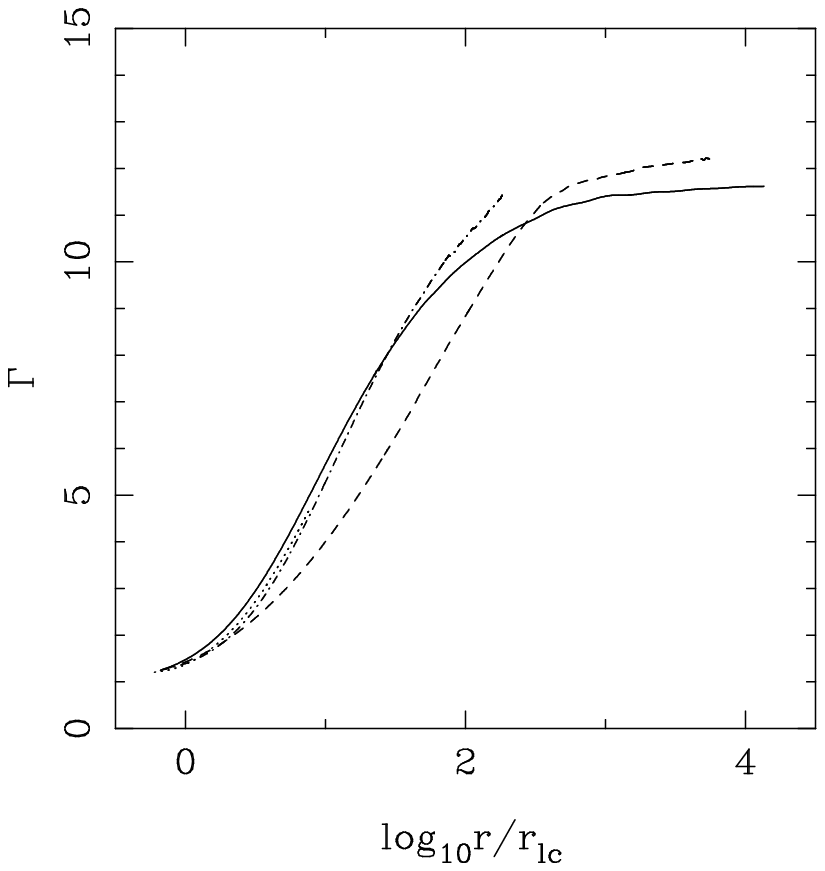}
\includegraphics[width=77mm,angle=-90]{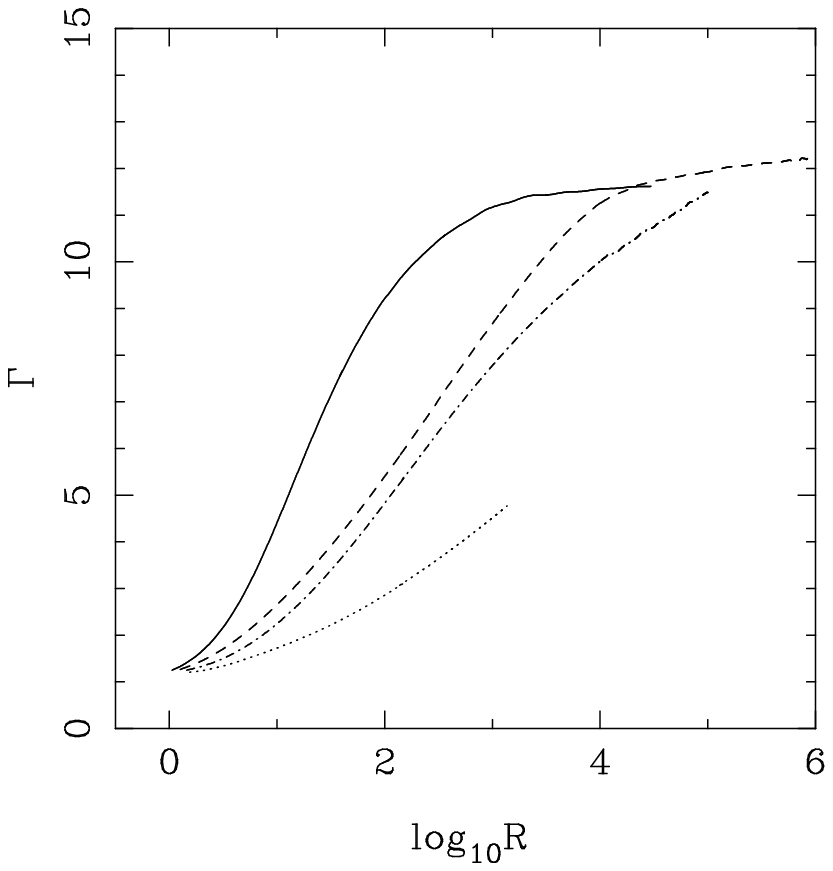}
\caption{Lorentz factor along the jet boundary as a function of 
the cylindrical radius $r$ (left panel) and the spherical radius
$R$ (right panel): model A2 -- solid line; model B2 -- dashed
line; model C2 -- dash-dotted line; model D2 -- dotted line. 
Note that, although $r_0$ and $r_{lc}$ differ from model to model,
$R_0=1$ in all cases.
}
\label{lor}
\end{figure*}

The left and middle panels of Fig.~\ref{sigma} show the evolution of
$\enint$, $\Gamma\sigma$ and $\Gamma$ along selected magnetic surfaces
for models C1 and C2.  For model C1 this flux surface is in the middle
part of the jet, where the flow accelerates most rapidly; it encloses
$\sim 1/3$ of the total magnetic flux in the jet. For model C2 this
surface is near the jet boundary, enclosing $\sim 5/6$ of the total
magnetic flux in the jet.  One can see that $\enint$ remains very nearly
constant on the surfaces, indicating that the flow has reached a steady
state and that the computational errors that we have described above are
fairly small. The Lorentz factor at first grows linearly with
cylindrical radius but then enters an extended domain of logarithmic
growth. 
The linear behaviour was previously found in the magnetically
dominated regime of self-similar solutions \citep[e.g.][]{VK03a},
whereas the logarithmic behaviour was shown to characterize the
acceleration in the asymptotic matter-dominated zone
\citep[e.g.][]{BL94}. The range of Lorentz factors in the solutions
derived in this paper is evidently too narrow to allow us to probe the
linear growth regime, but we expect that this could be done in our
forthcoming paper where we consider higher-$\Gamma_\infty$ flows.

The magnetization function $\sigma$ eventually
becomes less than 1, signaling a transition to the matter-dominated
regime. The right panel of Fig.~\ref{sigma} shows the evolution of
$\enint$, $\Gamma\sigma$ and $\Gamma$ along the magnetic flux surface
of model A2 that again encloses $\sim 5/6$ of the total magnetic flux
in the jet.  
This
conical jet also exhibits a very effective initial acceleration and a
transition to a matter-dominated regime.  In this case the growth of
the Lorentz factor saturates when it reaches $\Gamma\simeq 10$, a
value that corresponds to an acceleration efficiency $\Gamma/\enint$
of $77\%$.  Although the setup of our conical jet model most closely
evokes the conical flow geometries that have in previous works
produced very inefficient accelerations (see Section~\ref{theory}),
the results displayed in Fig.~\ref{sigma} demonstrate that this case
is not inherently different from the other ones. We discuss the
reasons for this in Section~\ref{theory}.
It is also worth noting that, while our choice of flux surfaces in
Fig.~\ref{sigma} was arbitrary, the behaviour along these surfaces is
fairly representative. To demonstrate this we show in Fig.~\ref{cross}  
the variation of the same parameters {\em across} the jets at large 
distances from the inlet. On can see that the jets are matter-dominated
throughout the entier cross-section.

Fig.~\ref{lor} compares the growth rates of $\Gamma$ in models
A2--D2. The numerical errors in these models are less restrictive than
in models A1--D1 and make it possible to extend the simulations to
larger spatial scales. Each of the plotted curves corresponds to the
magnetic surface near the jet boundary that encloses $\sim 5/6$ of the
total magnetic flux. The left panel shows $\Gamma$ as a function of
the cylindrical radius normalized by the light-cylinder radius.  The
most interesting feature of this figure is the very similar growth of
$\Gamma$ for all models. In fact, up to $r\sim 10-50\, r_{lc}$ the
curves for models A2, C2 and D2 are almost identical. In model B2 the
Lorentz factor increases somewhat more slowly. Further inspection
reveals another anomaly of model B2 -- in contrast to the C2 and D2
cases, where the highest Lorentz factor is found at the jet boundary,
the fastest acceleration in the B2 solution occurs somewhat off the
boundary.  The reason for these anomalies is not clear but it may have
something to do with the curvature of magnetic field lines -- given
the lower value of the power-law index $a$, model B2 retains a higher
curvature at larger radii than models C2 and D2.  The reason why the
model D curve is significantly shorter than the other is that the
strong collimation of the jet rapidly renders the computation
prohibitively expensive in this case.

Since more rapidly collimated jets reach the same cylindrical radius
at a larger distance from the source, the similar growth rates of the
Lorentz factor with cylindrical radius imply a faster growth with
spherical radius for less collimated jets. This is exactly what we see
in the right panel of Fig.~\ref{lor} -- the conical jet of the A2
model reaches a Lorentz factor of 10 at a distance from the origin
that is almost 100 times shorter than that of the paraboloidal jet of
model C2.

Fig.~\ref{ahu-b} compares the magnitudes of the different magnetic
field components in models A2--C2 near the far end of the jet
($\eta=10^3$). At this distance the jet radius is almost $10^3$ larger
than the light-cylinder radius and one would expect the azimuthal
component of the magnetic field to dominate. Indeed, this is what is
observed. On these scales the light cylinder is no longer resolved on
the computational grid, which explains why the azimuthal field
component exceeds the poloidal components even near the symmetry
axis. The fact that $B^{\hat{\phi}}$ does not vanish shows that the
solution develops a core of high electric current density, and this
core is also unresolved in our solution.  The development of an axial
line current in model C2 is shown in Fig.~\ref{cu-core}; very similar
results are found also for the other models.

\begin{figure*}
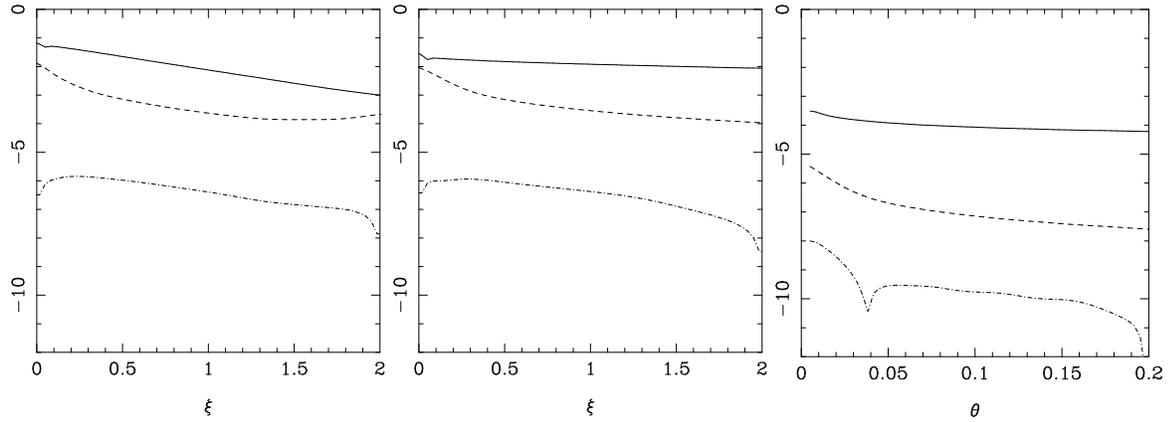

\includegraphics[width=55mm,angle=-90]{figures/c-bac.eps}
\includegraphics[width=55mm,angle=-90]{figures/cu-bac.eps}
\includegraphics[width=55mm,angle=-90]{figures/au-bac.eps}
\caption{Variation of the magnetic field components across the
jet at $\eta=6\times10^3$ in models C1 (left panel), C2 (middle
panel) and A2 (right panel): $\log_{10}|B^{\hat{\phi}}|$ --
solid line, $\log_{10}B^{\hat{\eta}}$ -- dashed line,
$\log_{10}|B^{\hat{\xi}}|$ -- dash-dotted line.
}
\label{ahu-b}
\end{figure*}

\begin{figure*}
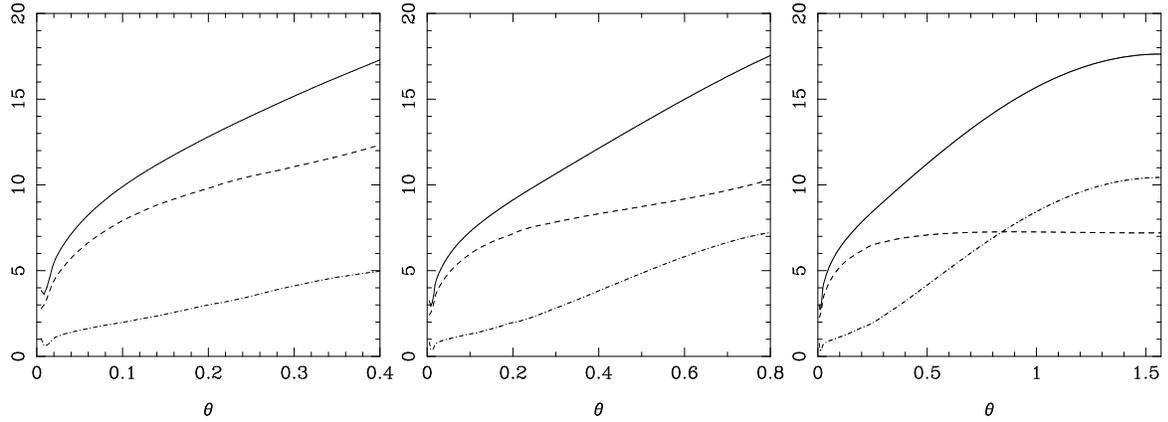

\includegraphics[width=55mm,angle=-90]{figures/a04-cross.eps}
\includegraphics[width=55mm,angle=-90]{figures/a08-cross.eps}
\includegraphics[width=55mm,angle=-90]{figures/api-cross.eps}
\caption{Distribution of $\enint$ (solid line), $\Gamma$ (dashed line) and
$\Gamma\sigma$ (dash-dotted line) across the jet for the cases of a
conical boundary with an opening
half-angle $\theta_c=0.4$~(left panel), 0.8~(middle panel), 
and $\pi/2$~(right panel, an unconfined flow) at $\eta=2\times10^3$. 
These are to be compared with the results for $\theta_c=0.2$ shown in 
the right panel of Fig.~\ref{cross}.
}
\label{crossn}
\end{figure*}

\begin{figure*}
\includegraphics[width=55mm,angle=-90]{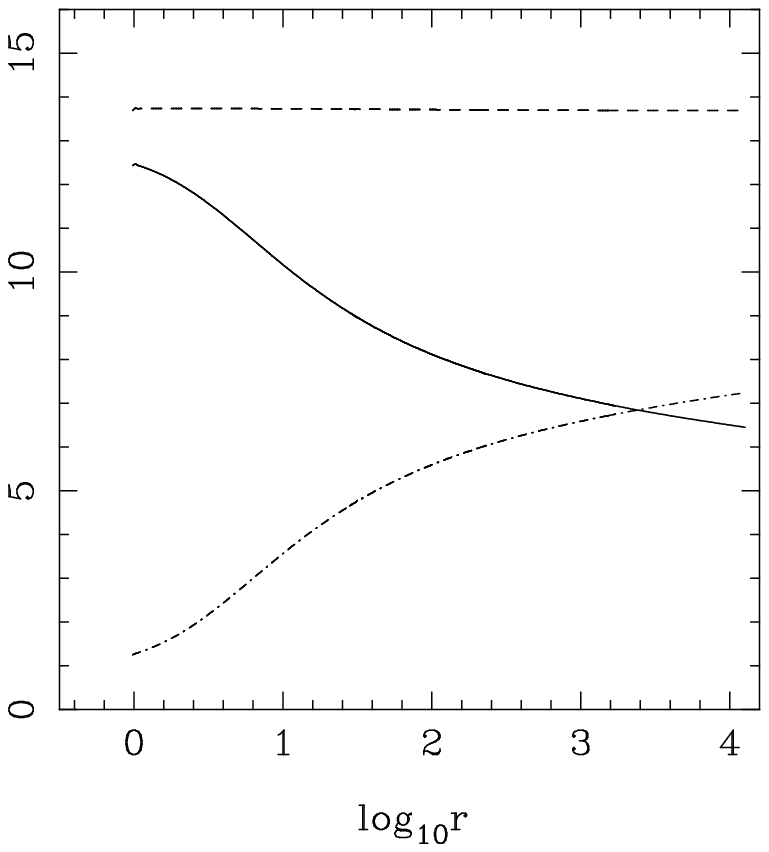}
\includegraphics[width=55mm,angle=-90]{figures/api-mflux.eps}
\caption{{\it Left panel:} Variation of   
$\Gamma\sigma$  (solid line), $\enint$  (dashed line) and
$\Gamma$ (dash-dotted line) along a magnetic field line for the case of
a conical boundary with $\theta_c=\pi/2$ (i.e. an unconfined flow). This
is to be compared with the results for $\theta_c=0.2$ shown in 
the right panel of Fig.~\ref{sigma}.
{\it Right panel:}
Evolution of the magnetic flux distribution across the unconfined jet. 
The solid line corresponds to $\eta=1$, the dashed line to
$\eta=30$, the dash-dotted line to $\eta=3\times10^2$, the dotted line
to $\eta=3\times10^3$ and the dash--triply-dotted line to
$\eta=3\times10^4$. This is to be compared with the results for 
$\theta_c=0.2$ shown in the right panel of Fig.~\ref{mflux}. 
}
\label{pwind}
\end{figure*}

In an attempt to clarify the role of the imposed wall on the
acceleration and collimation of the simulated jets, we performed
additional simulations of model A2, corresponding to
higher opening half-angles of the wall: $\theta_c=0.4, 0.8$, and
$\pi/2$, where the last case represents an unconfined outflow.\footnote{
Our original setup for this model corresponded to $\theta_c=0.2$.
We keep $\Delta\theta$ (the cell size in the polar direction) the same 
for all models by increasing the number of cells for larger $\theta_c$.}
The results are presented in Figs.~\ref{crossn} and~\ref{pwind}, which
should be compared with the corresponding plots in the right panels of
Figs.~\ref{mflux}--\ref{cross}. Not surprisingly, we find that the
efficiency of magnetic acceleration is lower for weaker external
collimation. This is particularly apparent in 
the unconfined case ($\theta_c=\pi/2$), where the flow remains
Poynting-dominated in the equatorial sector at all
distances. Nevertheless, even in the equatorial region of this extreme
case the acceleration efficiency $\Gamma_\infty/\enint$ is $\sim 40\%$. 
Furthermore, in the polar zone the efficiency is high, similarly to the
small-$\theta_c$
case, with all models exhibiting a transition to the matter-dominated regime.
In this sector self-collimation is strong as well (see Fig.~\ref{pwind}).
The behaviour of our unconfined jet model is consistent with the results
of \citet{Li96b}, who showed (for both the nonrelativistic and modestly
relativistic regimes) that a collimated axial jet can form from an
initially spherical MHD wind. Those results were, however, derived on
the assumption that the magnetic field is purely azimuthal and the
velocity field purely poloidal from the start. The good collimation
exhibited by all our jet models in the vicinity of the axis can be
attributed to the action of magnetic hoop stresses associated with a poloidal
current flow, as discussed in \S~\ref{theory}.

\section{Discussion}
\label{discussion}

\subsection{Theoretical aspects of the problem}
\label{theory}

Over the years there have been persistent doubts in the literature
regarding the ability of magnetic forces to accelerate flows to
relativistic speeds. In particular, several published studies have
concluded that MHD acceleration of relativistic flows is inherently
inefficient. This conclusion, however, is erroneous and can be
attributed to the adoption of a conical (split-monopole) flow geometry
in these studies. For example, in the work of \citet{M69} a simplified
conical geometry was used in which the full system of relativistic MHD
equations was not satisfied, whereas the results of \citet{BKR98} were
based on a perturbative analysis around a quasi-conical flow. The
conical flow geometry is unfavorable for acceleration for the
following reason. Well outside the light cylinder, where $r\Omega \gg
v^{\hat{\phi}}$ and $v\simeq c$, equations~(\ref{v_xi_b})
and~(\ref{v_phi_b}) imply
\begin{equation}
r B^{\hat{\phi}}=-\frac{1}{c} \Omega B_p r^2\, .
\end{equation} 
\noindent
From this equation and equation~(\ref{I}) one finds that
\begin{equation}\label{I-Bp} 
I=-\frac{1}{2} \Omega B_p r^2\, ,
\end{equation} 
\noindent
where $B_p$ is the magnitude of the poloidal magnetic field.  If the
magnetic surfaces are conical then $B_p \propto r^{-2}$, and thus the
poloidal electric current flows parallel to the magnetic field
lines. In this case the component of the Lorentz force along the
poloidal magnetic field lines, $(1/c) \vpr{j_p}{B_\phi}$, simply
vanishes. More general treatments of the problem, based on exact
semi-analytic solutions for axisymmetric, highly magnetized, steady
outflows under the assumption of radial self-similarity
\citep{LCB92,VK03a,VK03b,VK04}, have demonstrated that magnetic
acceleration in non-conical geometries can be quite efficient,
typically resulting in a rough asymptotic equipartition between the
Poynting and matter energy fluxes. A similar conclusion was reached on
the basis of a perturbative analysis around a {\it parabolic}\/ flow
\citep{BN06}. These results have indicated that the correct paradigm
should, in fact, be that magnetic acceleration is generally a rather
efficient mechanism for producing relativistic flows.

In this paper we have for the first time verified the proposed
paradigm by means of numerical simulations of highly magnetized,
relativistic flows. We have focused on the parameter regime that is
most relevant to AGN jets. In a future paper (Komissarov et al., in
preparation) we will present additional simulations that will
demonstrate that this paradigm also applies to flows with terminal
Lorentz factors that are as high as those inferred in gamma-ray burst
sources.

\begin{figure}
\includegraphics[width=77mm,angle=-90]{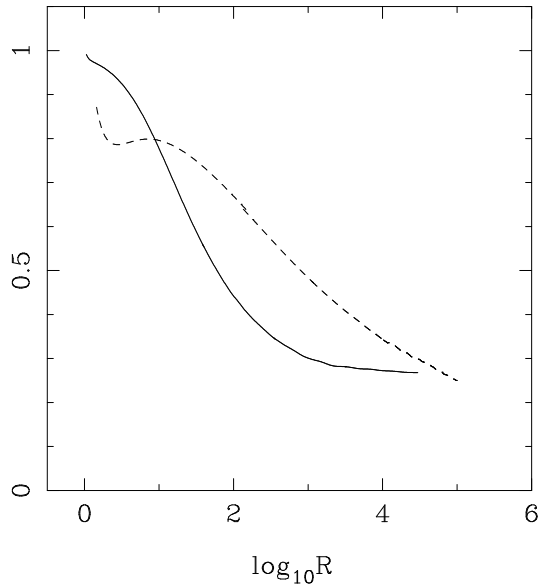}
\caption{Evolution of the function ${\cal S}=\pi B_p r^2/\Psi$ along 
the magnetic surfaces of models A2 (solid line) and C2 (dashed line).  
}
\label{ahu-mu}
\end{figure}

One of the interesting outcomes of this study is the highly effective
acceleration even in the case where the shape of the outer boundary is
conical. Although the acceleration efficiency in conical steady flows is
-- as explained above -- tiny, our results show that magnetic surfaces
of conical jets are not conical but rather paraboloidal (see
Figs.~\ref{lines} and~\ref{mflux}), so $B_p r^{2}$ is not
constant. Fig.~\ref{ahu-mu} shows the evolution of the function
\begin{equation}\label{S}
   {\cal S} = \frac{\pi r^2 B_p}{\Psi}
= \frac{\pi r^2 B_p}{\int \spr{B_p}{dS}}
\end{equation}
along a typical magnetic surface for models C2 and A2. It is seen that
in both cases ${\cal S}$ undergoes a significant decrease with
distance from the source. In fact, this decrease is faster in the
conical-boundary model, which is reflected in the more rapid
acceleration in this case (see Fig.~\ref{lor}). The direct
relationship between the function ${\cal S}$ and the acceleration
efficiency can be readily shown by combining
equations~(\ref{sigma-def}) and~(\ref{I-Bp}) to obtain
\begin{eqnarray}
\Gamma\sigma = (\Psi \Omega^2/ 4 \pi^2 \massint c^3)\, {\cal S}
\propto {\cal S}\, .  \nonumber
\end{eqnarray}

\begin{figure*}
\includegraphics[width=55mm,angle=-90]{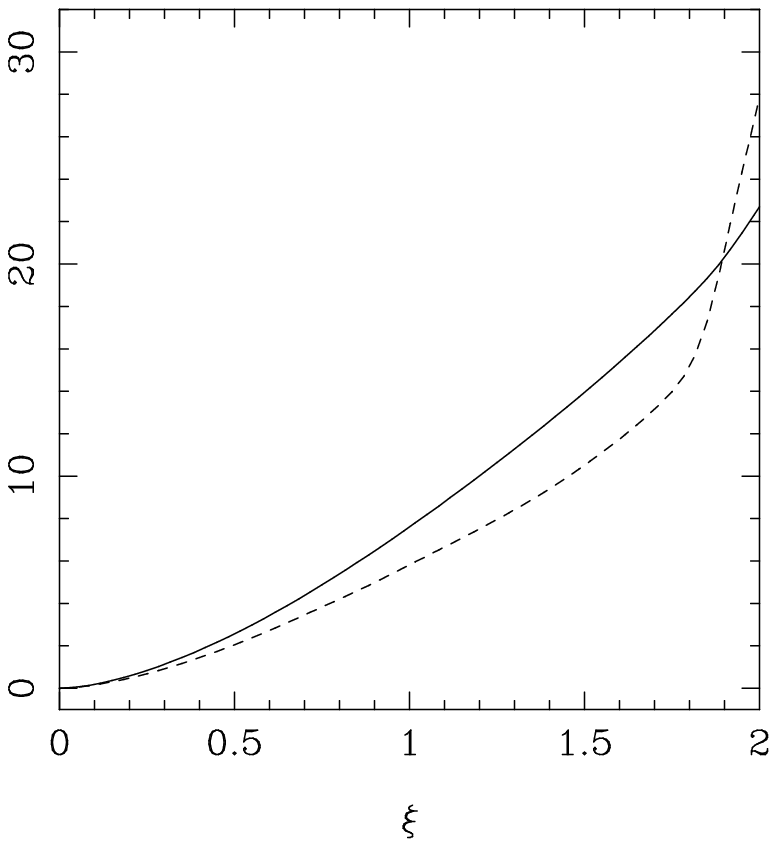}
\includegraphics[width=55mm,angle=-90]{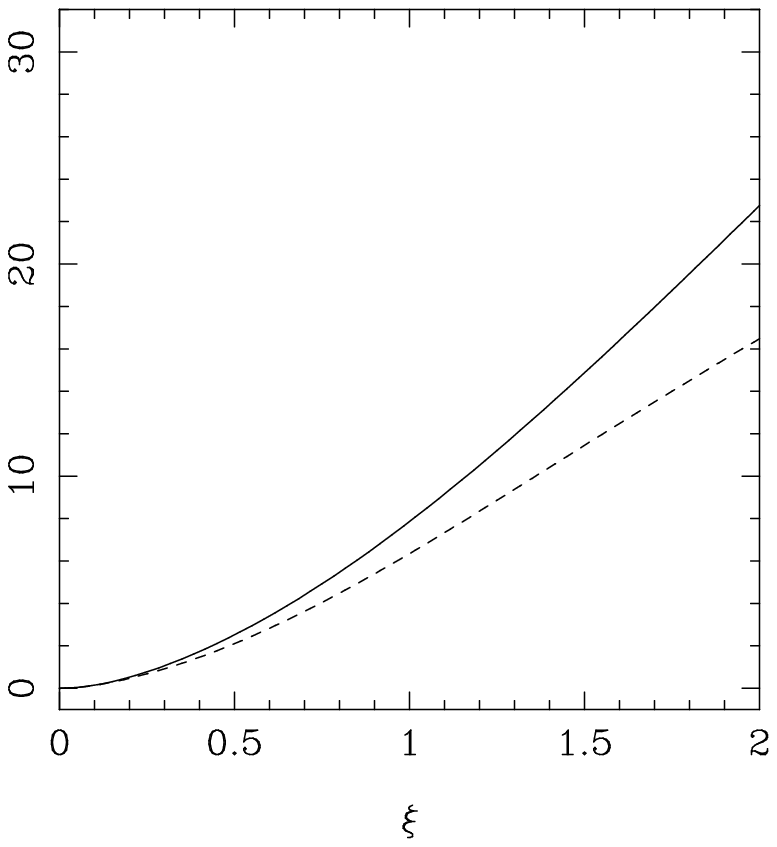}
\includegraphics[width=55mm,angle=-90]{figures/au-bpr2.eps}
\caption{Variation of $\Psi$ (solid line) and of $r^2 B_p$ (dashed
line) across the jet for models C1 (left panel), C2 (middle panel),
and A2 (right panel) at z=30.  Note that, for $z=30$,
$\xi=r/30^{1/a}\propto r$. The plotted curves therefore make it
possible to deduce the $r$-variation of the function ${\cal S}=\pi
(r^2 B_p) / \Psi$ at a constant $z$.  }
\label{neck1}
\end{figure*}

Since ${\cal S}$ depends on the shape of the flow, the latter relation
brings out the importance of the trans-field force balance and the
connection between acceleration and collimation.  If the poloidal
magnetic field is almost uniformly distributed across the jet then
${\cal S}\sim 1 $; this is the case near the inlet boundary. However,
due to the collimation, the poloidal magnetic flux becomes
concentrated near the rotation axis, forming a cylindrical core and
causing ${\cal S}$ to decrease with increasing $r$ (see
Fig.~\ref{neck1}).

Our jet solutions are characterized by a high magnetic-to-kinetic energy
conversion efficiency, but in the final states that we obtain the
kinetic and Poynting energy fluxes are still of comparable magnitude, as
previously found in the self-similar solutions \citep[e.g.][]{VK03a}. We
note in this connection that \citet{BL94} conjectured on the basis of
their asymptotic analysis that relativistic MHD outflows with a finite
total magnetic flux will tend to convert their entire magnetic energy
into kinetic energy at large distances from the origin. Although our
solutions are terminated at a finite distance and thus we cannot exclude
the possibility that eventually nearly all the magnetic energy will be
extracted, it appears that this is unlikely to occur on astrophysically
realistic scales. Furthermore, our discussion above suggests that the
reason given by \citet{BL94} for the limited conversion efficiency of
self-similar flows, namely, that it may be hard to meaningfully decrease
the quantity $\pi r^2 B_p$ in a medium that contains an infinite
magnetic flux $\Psi$, is not correct, since the relevant quantity is ${\cal
S}= \pi r^2 B_p/\Psi$ (equation~\ref{S}), which remains finite even in the
self-similar case.

Our results show efficient self-collimation, contrary to some claims in
the literature and certain seemingly similar investigations that reached
the opposite conclusion. 
As the term ``self collimation'' is sometimes misunderstood and confused 
with ``self confinement'', we start by clarifying its meaning. 
In general, a hydromagnetic jet cannot be ``self
confined'' for the simple reason that the ``outward'' effect of magnetic
pressure always overcomes the ``inward'' effect of magnetic tension in
MHD equilibria \citep[see e.g. Section 5.3 in][]{P79}. There must always
be an external medium that confines the system at its boundaries. (In
our case the geometry of the wall translates into an effective ambient
pressure distribution for any given flow problem; see Fig.~\ref{p_ext}
below.) It is, however, possible for a jet that carries an axial current
to have the magnetic hoop stress collimate the innermost
(current-carrying) streamlines relative to the outer regions of the
flow. This is what is generally meant by ``self collimation.'' The outer
streamlines may or may not be collimated, depending on the global
current flow and the confining pressure distribution
\citep[e.g.][]{BL94}. The behaviour of the innermost streamlines is
determined mostly by the axial current distribution. By showing that
these streamlines collimate faster than the imposed boundary, we were
able to demonstrate the action of the ``self collimation'' process in
our jet models,
including the limiting case of an unconfined flow (Fig.~\ref{pwind}). 
The different cases that we examined served to
illustrate how the transition from the inner region (small cylindrical
radius $R$) to the outer region (large $R$) occurs for different
effective confining pressure distributions.

Self-collimation in the super-Alfv\'enic regime of magnetized outflows
is the result of the $(1/c) \vpr{j_p}{B_\phi}$ force in the
trans-field direction. In relativistic flows, the effect of this force
is almost completely countered by the electric force, resulting in
slower collimation compared to the nonrelativistic case (where the
electric force is negligible). The asymptotic form of the trans-field
equation in the highly relativistic limit is
\begin{equation}\label{transf}
\frac{\Gamma^2 r}{{\cal R}} \approx \left[ \frac{ {\displaystyle
\left(\frac{2I}{\Omega B_p r^2} \right)^2 r
\vgrad{\ln\left|\frac{I}{\Gamma}\right|} } }{ {\displaystyle 1+\frac{4
\pi \rho u_p^2}{B_p^2} \frac{r_{lc}^2}{r^2}} }
-\Gamma^2\frac{r_{lc}^2}{r^2}\vgrad{r} \right] \!\cdot\!
\frac{\vgrad{\Psi}}{|\vgrad{\Psi}|}
\end{equation}
(see equation~16 in \citealp{V04}). From this equation it follows that
the radius of curvature of poloidal field lines is ${\cal R} \sim
\Gamma^2 r$. This fact led \cite{TB02} to propose a two-component
outflow model (central jet and surrounding disk wind) as a way of
explaining the collimation of relativistic jets. However, as the
self-similar solutions of \citet{VK04} as well as the present
simulations show, self-collimation is still possible.  In fact, it
remains possible in flows with even higher asymptotic Lorentz factors
\citep{VK03a}.  Although for $\Gamma\gg 1$ the collimation is indeed
slow, it is more efficient near the source, where the flow is not yet
highly relativistic.

\cite{B01} solved a similar problem (using time-dependent equations near
the central source and steady-state equations further out). Although the
setup in that paper is similar to our case A2, the conclusions are
different (inefficient collimation and therefore less efficient
acceleration compared to our solution).  There is, however, an important
difference in the two setups. In Bogovalov's (2001) paper the poloidal
velocity at the inlet boundary is $v_{p_{0}}\approx 0.87\, c$,
corresponding to a Lorentz factor (including the azimuthal velocity)
significantly higher than in our simulations.  As explained above, a
high Lorentz factor leads to a large ${\cal R}/r$ ($\sim \Gamma^2$).
Another difference between the two works is that we are able to follow
the flow to larger distances and hence to a smaller-$\sigma$ regime: in
fact, in some cases (A2 and B2) our solutions extend all the way to the
asymptotic regime, where the acceleration ceases and the Lorentz factor
saturates to a constant value.  Related to the above discussion is the
fact that the mass and magnetic flux in our jet solutions are not
``uncomfortably low'' (as they were acknowledged to be in Bogovalov's
2001 solution; see \citealt{TB02}). In our solutions all the magnetic
flux and all the outflowing mass are effectively collimated.\footnote{
The mass-loss rate between the axis and a particular field line
$\Psi=$const is $\dot{M}=2\int_0^\Psi \massint(\Psi) d\Psi$. Since
$\massint(\Psi)$ is practically constant (see Fig.~\ref{constants}),
$\dot{M}\propto \Psi$ and the distribution of mass-loss rate across the
jet can be deduced from the behaviour of $\Psi(r)$ in Fig.~\ref{neck1}.
}

The preceding discussion of the magnitude of the radius of curvature
implicitly assumes that ${\cal R}$ is positive. According to
equation~(\ref{transf}), the sign of ${\cal R}$ depends on the sign of
the three quantities $\vgrad{|I|}\cdot\vgrad{\Psi}$,
$\vgrad{\Gamma}\cdot\vgrad{\Psi}$ and $\vgrad{r}\cdot\vgrad{\Psi}$.
The term $\vgrad{r}\cdot\vgrad{\Psi}$ always corresponds to
decollimation and is important only in the matter-dominated flow
regime far from the source \citep{V04}. We can ignore this term in the
main acceleration region where the flow is still magnetically
dominated. The $\spr{j_p}{B_p}<0$ current-carrying regime (in which
$\vgrad{|I|}\cdot\vgrad{\Psi}>0$) contributes to positive ${\cal R}$
and thus to collimation, whereas the return-current regime
$\spr{j_p}{B_p}>0$ promotes decollimation \citep{O03}.  However, the
sign of ${\cal R}$ also depends on the gradient of $\Gamma$, a
manifestation of the electric force. It is possible to have ${\cal
R}>0$ even in the return-current regime provided that the Lorentz
factor decreases with increasing cylindrical radius
$(\vgrad{\Gamma}\cdot\vgrad{\Psi}<0$).  In this case the electric
force, which is directed toward the axis, dominates over the magnetic
force, which points away from the axis, leading to
collimation.\footnote{This behaviour was manifested also by the
return-current self-similar solution presented in \citet{VK03a}.}  The
net effect of the total electromagnetic force depends on the gradient
of $|I|/\Gamma$, and collimation is possible if this quantity
increases on moving across the field lines away from the polar axis
(see also \citealp{Li96b}).  In agreement with this analysis, the
positive value of ${\cal R}$ in our models A1--D1 (which contain a
current-carrying region near the axis and a return-current region near
the outer boundary) requires the Lorentz factor to decrease with
cylindrical radius (see e.g. Fig.~\ref{c1-2d}). The decrease in
$\Gamma$ as the outer wall is approached is consistent (by equation
\ref{sigma-def}) with the reduction in the electromagnetic
acceleration brought about by the imposition of the boundary condition
$\Omega=0$ at $\xi=\xi_j$ (see equation~\ref{Omega}).

To summarize the discussion on the collimation efficiency of
relativistic jets, we have argued that collimation {\it is}\/ possible
in accelerating flows where the Lorentz factor ranges from
$\Gamma_0\sim 1$ near the source to a high {\it asymptotic}\/
(subscript $\infty$) value $\Gamma_\infty$. Our choice of $v_{p_{0}}$
($=0.5\, c$) at the inlet boundary and of the initial value of
$v^{\hat{\eta}}$ for the funnel plasma ($=0.7\, c$) allows the flow to relax
to a collimated steady state with a high $\Gamma_\infty$.  The sign of
the curvature radius is positive even in the return-current regime
because the Lorentz factor decreases sufficiently rapidly with $r$
across the jet.

\begin{figure*}
\includegraphics[width=55mm,angle=-90]{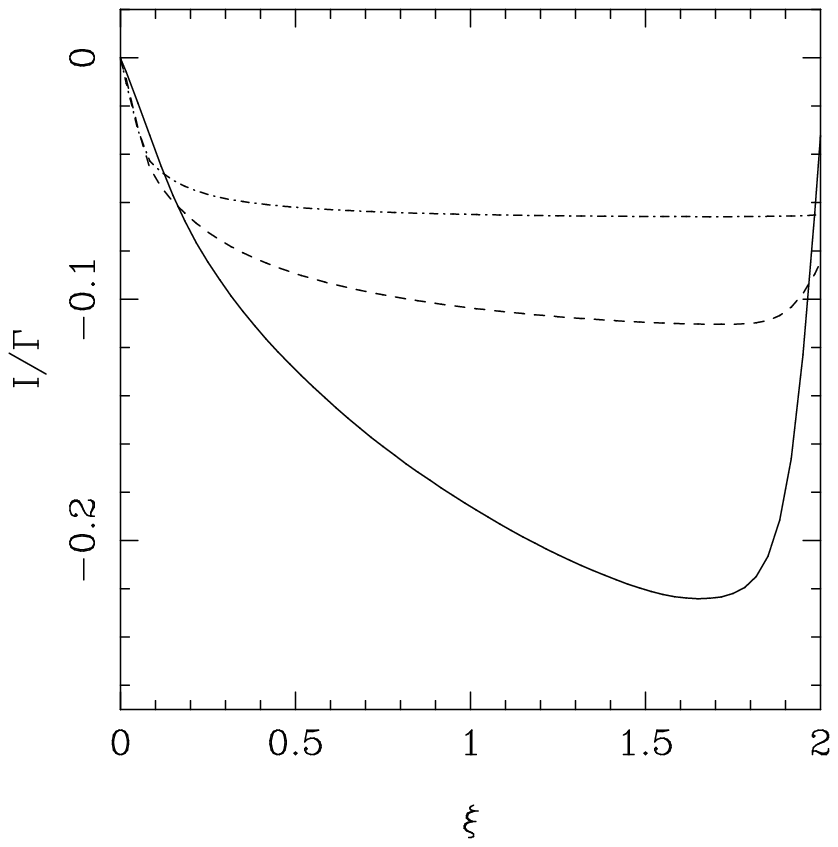}
\includegraphics[width=55mm,angle=-90]{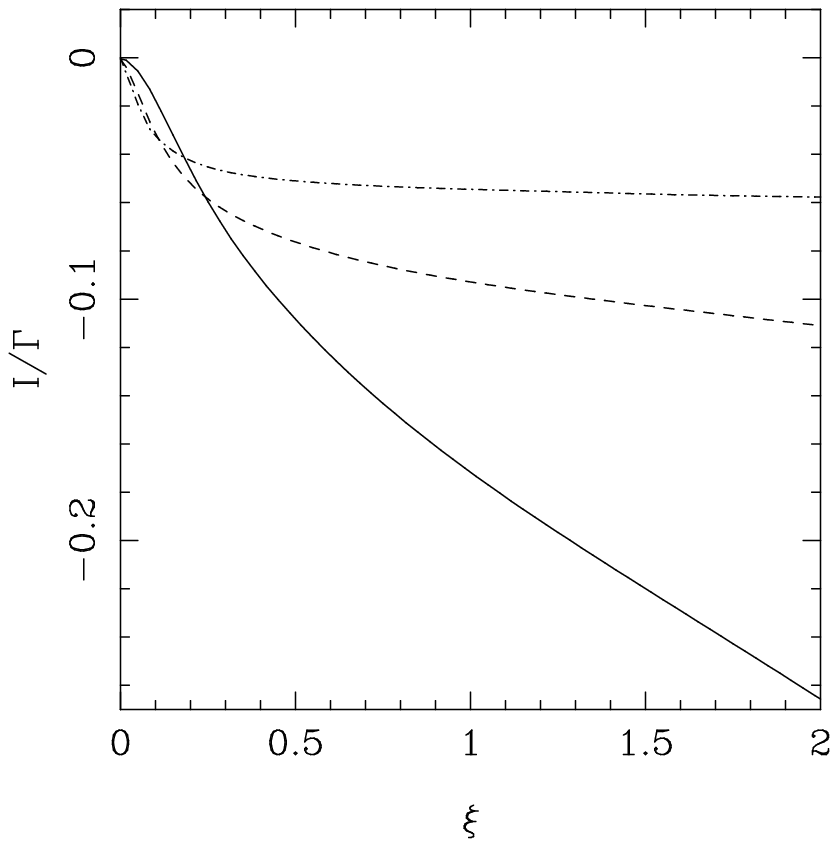}
\includegraphics[width=55mm,angle=-90]{figures/au-par1.eps}
\caption{Distribution of $I/\Gamma$ across the computed jets.
{\it Left panel}: Model C1 at $\eta=80$ (solid line), $\eta=800$ (dashed line)
and $\eta=8000$ (dash-dotted line); {\it Middle panel}: Model C2 at
$\eta=80$ (solid line), $\eta=800$ (dashed line) and $\eta=8000$
(dash-dotted line); {\it Right panel}: Model A2 at $\eta=20$ (solid
line), $\eta=200$ (dashed line) and $\eta=20000$ (dash-dotted line).
}
\label{par1}
\end{figure*}

According to the asymptotic analysis of \cite{HN89} and its
relativistic generalization by \cite{CLB91}, the formation of a
cylindrical core is the only way to have a
non-singular current near
the axis, i.e. $I_\infty(\Psi=0)= 0$. Although our numerical
results do not include the far-asymptotic state, the ``solvability
condition'' $I_\infty(\Psi)/\Gamma_\infty(\Psi)=$const (a direct
consequence of equation~\ref{transf} in the limit $r/{\cal R}
\rightarrow 0^+$) is roughly satisfied; see Fig.~\ref{par1}.

The profile of $\Omega(\xi)$ on the inlet boundary is used in our
simulations as a proxy for the current distribution (see
equation~\ref{I-Bp}).  However, in the cases A1--D1, where $\Omega$
vanishes on the outer boundary, $I$ nevertheless remains finite there.
(Note that equation~\ref{I-Bp} holds only for $r\Omega \gg
v^{\hat{\phi}}$, which is not satisfied when $\Omega$ vanishes.)  This
is a reflection of a general property of ideal MHD flows: in the
super-Alfv\'enic regime the azimuthal magnetic field cannot
vanish\footnote{ As explained in \cite{V04}, the invariance of
$B^2-E^2$ yields $B_\phi^2/B_p^2>(r^2/r_{lc}^2)-1$. In highly
magnetized flows the Alfv\'en surface almost coincides with the light
surface, so $(r^2/r_{lc}^2)-1>0$ in the super-Alfv\'enic regime.}  and
the current lines close in a current sheet. In this respect our choice
of an outer ``wall'' captures a basic physical aspect of a real
boundary between a jet and an unmagnetized environment.

Since we have constructed our numerical jet models using time-dependent 
simulations, we have partly addressed the issue of stability to axisymmetric
perturbations.  Our results indicate that, within our model setup, the
jets are stable. Although the imposed solid outer boundary could 
have some stabilizing effect even on these modes, the extent of 
this influence is unclear, although in any case we do not expect the
outer wall alone to prevent the growth of the pinch mode in the jet interior.  
Linear stability analyses of ``Z-pinch'' equilibria (magnetostatic cylindrical 
configurations with a weak uniform poloidal field and a strong toroidal
field) have been used to argue that the kink mode would destroy the
concentric configurations  of astrophysical jets \citep{Beg98}. While
our simulations cannot address the possible effect of nonaxisymmetric
modes, we nevertheless note that, in contrast to ``Z-pinch'' configurations, 
our jets show strong transverse gradients of poloidal velocity and 
magnetic field, which may be stabilizing factors \citep[e.g.][]{AZKB}. 

\subsection{Application to AGN Jets}
\label{application}

The initial energy-to-mass flux ratio of jets in our simulations
yields an upper limit on the terminal Lorentz factor $\Gamma_\infty=
\enint \leq 16$.  This is consistent with the mean values inferred in
AGN jets (see Section~\ref{introduction}). In order to make further
comparisons of our numerical models with observations we need to
select suitable dimensional scales. The key scale in the problem of
magnetic acceleration is the light cylinder (or the Alfv\'en surface)
radius, $r_{lc}$.  If the jets are launched by a rapidly rotating
black hole in the center of an AGN then
\begin{eqnarray}
\nonumber r_{lc}\simeq 4 r_g = 6\times 10^{13} ( {M}/{10^8 \,
M_{\sun}} ) \mbox{ cm}\, ,
\end{eqnarray}
where $r_g\equiv GM/c^2$. In this estimate we assume that the angular
velocity of the magnetic field lines is half of that of a maximally
rotating (rotation parameter $a\simeq1.0$) black hole.  According to the
results shown in Fig.~\ref{sigma}, the jets enter the matter-dominated
regime, where an equipartition between Poynting and matter energy fluxes
($\sigma \simeq 1$) is established, at a cylindrical radius
\begin{eqnarray}
\nonumber r_{eq} \simeq 30\, r_{lc} \simeq 2\times 10^{15} ({M}/{10^8
M_{\sun}}) \mbox{ cm}\, ,
\end{eqnarray}
more or less independently of the details of jet collimation.  The
corresponding distance from the black hole is
\begin{eqnarray}
\nonumber
   R_{eq} \simeq 2\times 10^{16} \left( \frac{M}{10^8 M_{\sun}} \right)
   \left( \frac{\Theta_j}{0.1}\right)^{-1} \mbox{ cm}\, ,
\end{eqnarray}
where $\Theta_j$ is the jet opening half-angle.  If blazar flux
variability is associated with the propagation of strong shocks within
the jet then we can expect this behaviour to originate on scales $\ga
R_{eq}$. When our simulated jets reach $R \simeq 10 \, R_{eq}$, their
characteristic Lorentz factor becomes $\sim 10$. These properties of
the extended magnetic acceleration region are in very good agreement
with the observational inferences summarized in
Section~\ref{introduction} \citep[in particular, the lack of
bulk-Comptonization spectral features in blazar jets; see][]{S05}.

Taking the characteristic initial radius poloidal magnetic field of a
black hole-launched jet to be $r_0=r_g$ and $B_0=10^5$G, respectively,
the mass-loss rate and total luminosity of the model jets scale as
\begin{eqnarray}
\nonumber
\dot{M}} = 10^{-2}
\left (\frac{\massint\Psi}{0.2}\right )
\left( \frac{B_0}{10^5\, \mbox{G}} \right)^2
\left( \frac{r_0}{r_g} \right)^{2}
\left( \frac{M}{10^8\, M_{\sun}} \right)^2 {M_{\sun}\, \mbox{yr}^{-1}
\end{eqnarray}
and
\begin{eqnarray}
\nonumber
\dot{\cal E}= 6 \times 10^{45}
\left( \frac{\enint}{10} \right)
\left( \frac{\dot{M}}{10^{-2}\, M_{\sun}\, \mbox{yr}^{-1}} \right)
\mbox{ ergs s}^{-1} \, ,
\end{eqnarray}
respectively, where $\massint$, $\Psi$ and $\enint$ are the mean values
of the dimensionless flow constants shown in Fig.~\ref{constants}.

From the theory of black-hole magnetospheres \citep[e.g.][]{BZ77} it
follows that, at their base, black-hole jets are highly magnetically
dominated, so that the energy per particle greatly exceeds the Lorentz
factor inferred from observations of AGN jets. This difficulty can be
overcome if there are other ways of injecting particles into AGN jets
in addition to pair cascades. It is conceivable that a sufficient
supply of particles is provided by the winds of stars that lie in the
paths of the jets as they make their way out of the galactic nuclei
\citep[e.g.][]{K94,HB06}. In fact, the injection rate could be high
enough to explain the observed deceleration of weak FR-I jets down to
subrelativistic speeds.  
In addition some mass is unavoidably entrained through the interface 
between the jet and the confining external medium via diffusive
processes.  Another possibility is that the speed of AGN jets
is controlled by the dissipation at oblique shocks developing 
as a result of an interaction with an unsteady external flow or due to
various instabilities \citep{McK06}. Compton-drag interaction with the ambient 
radiation field  may also be a factor \citep[e.g.][]{MK89}.

The problem of low initial mass loading might be circumvented if the
bulk of the relativistic outflow component in fact originates in the
nuclear accretion disk \citep[see, e.g.][]{GA97,LOP99}, which is the
scenario adopted in the semi-analytic models of \citet{VK04}. Although
the chosen distributions of angular velocity, magnetic field and density
at the inlet boundary do not formally match a Keplerian disk, our
solutions can be interpreted in the context of disk-driven
outflows. Taking $\Omega_{\rm K} (r_0)$ to be the Keplerian angular
velocity at the reference distance $r_0$, we find that
\begin{eqnarray}
\nonumber
r_{lc}=5 \times 10^{14} \left( \frac{M}{10^8 M_{\sun}} \right) 
\left( \frac{r_0}{10 \, r_g} \right)^{3/2} 
\left[ \frac{\Omega_{\rm K} (r_0)}{\Omega} \right ] \mbox{ cm}
\end{eqnarray}
and that $\sigma \approx 1$ is reached at a distance
\begin{eqnarray}
\nonumber
R_{eq} \simeq 10^{17} 
\left( \frac{r_{lc}}{5 \times 10^{14} \mbox{ cm}} \right)
\left( \frac{\Theta_j}{0.1}\right)^{-1} \mbox{ cm}\, .
\end{eqnarray}
The mass-loss rate and jet power are
\begin{eqnarray}
\nonumber
\dot{M}} = 10^{-2}
\left (\frac{\massint\Psi}{0.2}\right ) 
\left( \frac{B_0}{10^4\, \mbox{G}} \right)^2 
\left( \frac{r_0}{10\, r_g} \right)^{2}
\left( \frac{M}{10^8\, M_{\sun}} \right)^2 {M_{\sun}\, \mbox{yr}^{-1}
\end{eqnarray}
and
\begin{eqnarray}
\nonumber
\dot{\cal E}= 6 \times 10^{45}
\left( \frac{\enint}{10} \right)
\left( \frac{\dot{M}}{10^{-2}\, M_{\sun}\, \mbox{yr}^{-1}} \right)
\mbox{ ergs s}^{-1} \,,
\end{eqnarray}
respectively. 
The striking differences in the velocity profile between simulated jets
with solid-body rotation and with differential rotation (see Figs.~\ref{c1-2d}
and~\ref{c2-2d} for models C1 and C2, respectively) suggests (taking
into account relativistic beaming effects) that it
might be possible to distinguish between jets launched directly from a
black hole and those that emanate from the surface of an accretion disk
in cases where the transverse structure of the jet can be resolved. One
would, however, need to verify that these differences remain noticeable
for more realistic disk rotation laws and surface field distributions.

\begin{figure}
\includegraphics[width=77mm,angle=-90]{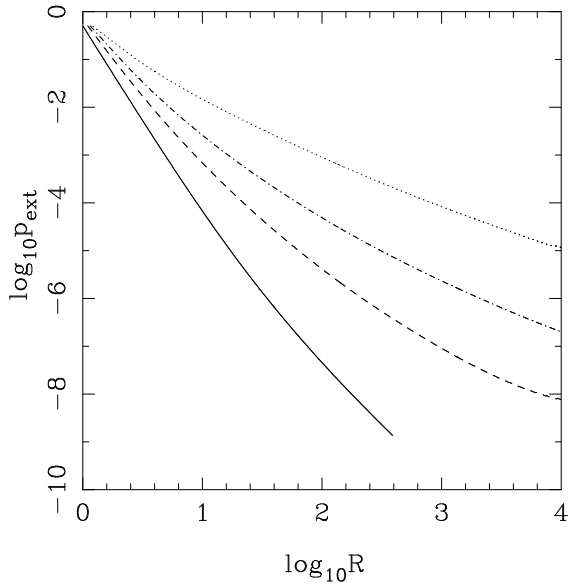}
\caption{ Ambient pressure distribution required for the collimation
of the computed jets. Model A2 -- solid line; model B2 -- dashed line;
model C2 -- dash-dotted line; model D2 -- dotted line.  }
\label{p_ext}
\end{figure}

The funnel shapes in our simulations were chosen merely for the
purpose of studying the effects of overall flow collimation on its
magnetic acceleration.  However, from our steady-state solutions we
can infer the effective external force (normal to the jet surface)
that is required to provide the collimation imposed by our choice of
the outer-boundary shape. AGN jets could be confined by a variety of
forces, including, for example, the thermal pressure of an ambient gas
distribution, the ram pressure of a wind from the outer regions of the
nuclear disk, and the stress of a magnetic field anchored in the disk
(and possibly embedded in a disk outflow). Fig.~\ref{p_ext} shows the
effective pressure deduced in this way ($p_{ext}=b^2/8\pi$) as a
function of spherical radius for models A2--D2.  Although none of the
curves is an exact power law of the form $p_{\rm ext} \propto
R^{-\alpha}$, it is nevertheless informative to calculate mean
power-law indices. We find $\alpha \approx 3.5$, 2, 1.6 and 1.1 for
models A2, B2, C2 and D2, respectively. The models are thus seen to
cover a wide range of behaviours. As expected, the more highly
collimated funnel geometries correspond to the less steeply declining
effective pressure distributions. The largest indices might correspond
to confinement by a wind. For example, in a spherical wind of
polytropic index 5/3, the thermal pressure scales as $R^{-10/3}$ and
the ram pressure as $R^{-5/2}$. Thus, a disk wind that assumes a
nearly spherical geometry as it propagates away from the disk surface
could effectively confine a relativistic jet with a nearly conical outer
boundary. 
We note in this connection that \citet{BL94} suggested that
collimated jets necessarily correspond to confining pressure
distributions with $\alpha \la 2$. Our results indicate that higher
values of the effective power-law index are also possible.

Although we have focused in this paper on AGN jets, it is interesting
to note that relativistic outflows with terminal Lorentz factors as
high as $\sim 10$ have also been inferred in Galactic X-ray binary
sources, which comprise both black holes and neutron stars
\citep[e.g.][]{F04}, and that arguments have been advanced in support
of the possibility that the mean Lorentz factors in these sources are
comparable to those estimated in AGNs \citep{MFN06}. The magnetic
acceleration mechanism discussed in this paper is also a likely
candidate for the driving of X-ray binary jets
\citep[e.g.][]{LPK03}. However, even if these jets are similar to
those in AGNs, as of now the latter remain the best targets for
observations that could test and constrain the model.


\section{Conclusion}
\label{conclusion}

This paper presents the results of special-relativistic, ideal-MHD
numerical simulations of AGN jets. The numerical code employed in
these simulations was specifically designed for this task. In contrast
to most previous numerical schemes that modeled relativistic MHD jets,
our code does not require a large artificial viscosity for numerical
stability and is well suited for studies of two-dimensional stationary
flows that are aligned with the computational grid. To avoid numerical
dissipation induced at the interface with an ambient medium, we have
simplified the calculation by taking the flow to be confined by a
solid wall. We took the shape of the wall to be either a paraboloid of
revolution or a cone and used corresponding elliptical or spherical
coordinates to optimize the resolution as well as the alignment of the
flow with the computational grid. In addition, we implemented a
grid-extension method that allowed us to follow the flow out to scales
$\sim 10^4 - 10^6$ times that of the inlet boundary, which was crucial
to our ability to study the inherently extended nature of MHD
acceleration to high Lorentz factors. To ensure the self-consistency
of this procedure, we derived the condition for the Mach cone of the
fast-magnetosonic waves to point outward at the boundary between a
given grid sector and the successive one, and we verified that this
condition is satisfied at each of the relevant sector interfaces.

Our carefully designed numerical scheme has enabled us to simulate, for
the first time, the magnetic acceleration and collimation of
relativistic jets to completion. In particular, we have found that
initially Poynting flux-dominated jets can be effectively accelerated to
high bulk Lorentz factors with an efficiency (defined as the ratio of
the final kinetic energy flux to the total energy flux) $\ga 50\%$. As
expected from previous semi-analytic (radially self-similar) solutions
for steady-state flows, the acceleration process is spatially
extended. We have found that our simulated jets invariably settle to a
steady state, which suggests (although we did not explore this issue
explicitly) that the resulting flow configurations are not inherently
unstable (at least not to axisymmetric perturbations -- although it is
conceivable that the imposed rigid wall has a stabilizing influence in
this regard -- and excluding by design any effects of a direct
interaction with an ambient medium). The properties of the derived final
configurations were found to be qualitatively very similar to those of
the self-similar AGN jet solutions of \citet{VK04} and to not depend
sensitively on either the imposed shape of the outer boundary or on the
distribution of the injected poloidal current at the inlet boundary. (We
explored boundaries with scalings, in cylindrical coordinates, ranging
from $z\propto r$ to $z\propto r^3$, and current distributions that
either closed within the volume of the jet or on its outer boundary.)

We provided a physical explanation of the basic acceleration process
and of the variations in the detailed behaviour among the different
flow configurations that we simulated. We argued that the robustness
of the acceleration process can be attributed to the fact that the
bulk of the outflow initially follows paraboloidal trajectories,
including the case of a conical outer boundary. We highlighted the
connection between the collimation of the flow, which is manifested in
the curved streamlines, and the acceleration process. The collimation
in the current-carrying regime is essentially due to magnetic hoop
stress associated with the azimuthal magnetic-field component
$B_\phi$. The collimation induces a reduction in the magnitude of $r^2
B_p$ (where $B_p$ is the poloidal field component) along the poloidal
streamlines, which corresponds to a decrease in the Poynting flux
along the flow and therefore results in acceleration (driven by the
gradient of the magnetic pressure associated with $B_\phi$). Previous
claims in the literature that magnetic acceleration of relativistic
flows is inefficient were all based on the assumption that the
streamlines have a split-monopole geometry (or very nearly so), which
is a singular case in which by fiat $r^2 B_p$ remains constant (or
close to a constant) along the flow.

Our solutions also revealed that the collimation efficiency of
relativistic jets can be high if they are accelerated from an initial
Lorentz factor $\Gamma_0 \sim 1$. We argued that published results in
the literature that claimed otherwise in fact had a significantly higher
$\Gamma_0$. Once the flow attains a high Lorentz factor the collimation
process slows down on account of the increased inertia and of the growth
of the electric force, which almost completely balances the transverse
magnetic force. Nevertheless, the current-carrying central region of our
simulated jets collimates much faster than the imposed boundaries and
attains a cylindrical shape by the time the terminal Lorentz factor is
attained, again in full agreement with the semi-analytic self-similar
solutions (which also assumed $\Gamma_0 \sim 1$). In simulated outflows
where the current returns through the jet we found that the flow is
effectively collimated also in the outer, return-current region (in this
case by the electric force, which dominates the transverse magnetic
force that acts to decollimate the flow in this regime). The efficient
electromagnetic collimation in all of our computed jet models is
evidently the reason why the presence of a rigid outer boundary does not
induce recollimation shocks in the outflow even for the most rapidly
converging wall shape ($r \propto z^3$).

In validating the basic features of the simplified semi-analytic
solutions, our numerical results go a long way toward establishing an
``MHD acceleration and collimation paradigm'' for relativistic
astrophysical jets. In this contribution we applied this model to AGN
jets, for which there is already significant observational evidence of
extended, $\ga 0.1\, {\rm pc}$-scale acceleration (possibly continuing
to $\sim 1-10\, {\rm pc}$) to Lorentz factors $\ga 10$. We
demonstrated that, for plausible physical parameters, our simulated
jets can reproduce these observations (see also \citealt{VK04}). We
noted that these results could potentially apply also to the jets
observed in Galactic X-ray binary sources. In a future publication we
will present simulation results for even higher initial magnetizations
that could be used to model the ultrarelativistic jets in gamma-ray
burst sources.

Future investigations will need to examine some of the complicating factors
left out of the present study. In particular, the simplifying wall
boundary condition will have to be 
replaced by the interaction of the outflow with its environment.
Furthermore, the questions of stability and energy dissipation will
have to be addressed.
In fact, given the potentially destructive role of the kink mode, fully
3D relativistic simulations will ultimately need to be carried out.  
In addition, the response of the jet to nonsteady conditions at the source
-- a potential cause of observed variability -- will be an interesting
topic for study.

\section*{Acknowledgments}
We thank Mitch Begelman and Anatoly Spitkovsky for helpful comments and
Jonathan McKinney for useful correspondence.
We also acknowledge extensive suggestions by the referee.
This research was funded by PPARC under the rolling grant
``Theoretical Astrophysics in Leeds'' (SSK and MVB).
NV acknowledges partial support 
by the European Social Fund and National Resources -- (EPEAEK II) PYTHAGORAS. 
AK was partially supported by a NASA Theoretical Astrophysics Program grant.
\appendix
\section{Elliptic Coordinates}
\label{appendix_1}

We assume that the jet boundary satisfies the power law 
\begin{equation}
  z \propto r^a\, ,
\label{a0}
\end{equation}
where $\{r,z\}$ are cylindrical coordinates. Then the condition that 
the jet boundary be a coordinate surface suggests we choose
\begin{equation}
 \xi=rz^{-1/a}
\label{a1}
\end{equation}
as one of the spatial coordinates. The other coordinate, $\eta$, is defined in 
such a way that the coordinate system becomes orthogonal. The orthogonality
condition 

\begin{equation}
 \vgrad{\xi}\cdot\vgrad{\eta}=0
\label{a2}
\end{equation}
leads to a PDE for $\eta$\, ,

\begin{equation}
\pder{\eta}{r}-\frac{1}{a}\frac{r}{z}\pder{\eta}{z} =0\, ,
\label{a3}
\end{equation}
which allows separable solutions. The requirement $\eta=0$ for $(r,z)=(0,0)$
leads to 
\begin{equation}
 \eta^2=\frac{r^2}{a} + z^2\, .
\label{a4}
\end{equation}
Thus, the $\eta$ coordinate lines are ellipses with semi-axes $\eta$ and
 $\sqrt{a}\eta$. The remaining spatial coordinate is the usual azimuthal angle
$\phi$.

Conversion from elliptical to cylindrical coordinates involves
solving a transcendental equation for $z$:

\begin{equation}
 z^2+\frac{\xi^2}{a} z^{2/a} -\eta^2 =0\, .  
\label{a5}
\end{equation}
In the general case this equation has no analytic solutions, but
for certain values of the power-law index $a$ it reduces to simpler equations:
\begin{eqnarray} 
a=3/2, \quad y^3+\frac{2}{3}\xi^2y^2-\eta^2=0, \text{where}
y=z^{3/2}\, ;\quad\\
a=1, \quad z^2+\xi^2 z^2 -\eta^2=0\, ; \text{
 }\qquad\qquad\qquad\qquad\quad\\ 
a=2, \quad z^2+\frac{\xi^2}{2}z-\eta^2=0\, ; \text{
 }\qquad\qquad\qquad\qquad\quad\\ 
a=3, \quad y^3+\frac{\xi^2}{3}y-\eta^2=0, \text{where}
y=z^{2/3}\, .\qquad\quad   
\end{eqnarray}   
 
The metric tensor in these coordinates is diagonal with components
\begin{equation}
g_{\xi\xi}= \frac{a^2 z^{2(1+a)/a}}{D}\, ,
\label{a6}
\end{equation}
\begin{equation}
g_{\eta\eta}=\frac{a^2\eta^2}{D}\,  ,
\label{a7}
\end{equation}
\begin{equation}
g_{\phi\phi}= r^2\, ,
\label{a8}
\end{equation}
\begin{equation}
g_{tt}= -1\, ,
\label{a9}
\end{equation}
where $D=a^2z^2+r^2$. Its determinant is 
\begin{equation}
g= -\frac{a^4r^2\eta^2z^{2(1+a)/a}}{D^2}\, . 
\label{a10}
\end{equation}
The non-vanishing derivatives of these components are 
\begin{equation}
g_{\xi\xi,\xi} = -\frac{1}{D^3} 2a^2 r z^{(3+2a)/a} 
     [2a^2z^2+r^2(1+a)]\, ,
\label{a11}
\end{equation}
\begin{equation}
g_{\xi\xi,\eta} = \frac{1}{D^3} 2a^5 \eta^3 z^{(2+2a)/a}\, ,
\label{a12}
\end{equation}
\begin{equation}
g_{\eta\eta,\xi}=\frac{1}{D^3} 2a^4(a-1) r z^{(1+2a)/a}\eta^2\, ,
\label{a13}
\end{equation}
\begin{equation}
g_{\eta\eta,\eta}=-\frac{1}{D^3} 2a^3 (a-1)^2 r^2 z^2\eta\, ,
\label{a14}
\end{equation}
\begin{equation}
g_{\phi\phi,\xi}=\frac{1}{D} 2a^2 r z^{(1+2a)/a}\, ,
\label{a15}
\end{equation}
\begin{equation}
g_{\phi\phi,\eta}=\frac{1}{D}2a r^2 \eta\, .
\label{a16}
\end{equation}

\section{Fast Magnetosonic waves}
\label{appendix_2}
Suppose that we study an axisymmetric magnetosonic disturbance whose
wavevector is
\begin{eqnarray}
\bmath{k}= k \left(\frac{\bmath{v}_p}{v_p} \cos \thetak
+ \hat{\phi}\times \frac{\bmath{v}_p}{v_p} \sin \thetak \right)
\,.
\end{eqnarray}
Its frequency is $\omega=k {\cal P}(\thetak)$, with ${\cal P}$ satisfying
\begin{eqnarray}
\label{dispersion}
\frac{B_p^2 c_s^2}{ 4 \pi {w}/{c^2}}
\left[\cos^2 \thetak
-\left(\frac{r}{r_{lc}}\cos \thetak +\frac{{\cal P}}{c} 
\frac{B_\phi}{B_p}\right)^2
-\frac{{\cal P}^2}{c^2} \right]
\nonumber \\
- \left(\Gamma {\cal P} - u_p \cos \thetak \right)^2
\left[ \frac{B^2-E^2}{4 \pi {w}/{c^2}} \left(1-\frac{c_s^2}{c^2}\right)
+ c_s^2  \right]
\nonumber \\
+\frac{\left(\Gamma {\cal P} - u_p \cos \thetak \right)^4}{1-{\cal P}^2/c^2}
\left(1-\frac{c_s^2}{c^2}\right)
=0
\end{eqnarray}
(see Appendix C in \citealp{VK03a}). The disturbance travels with a group velocity
\begin{eqnarray}
\bmath{v}_g=\nabla_{\bmath{k}} \omega =
\frac{\bmath{k}}{k} {\cal P} + \hat{\phi}\times \frac{\bmath{k}}{k} {\cal P}'
\,,
\end{eqnarray}
where ${\cal P}' \equiv d{\cal P}/d\thetak$.
The group velocity makes an angle $\thetag$ with respect to the poloidal
flow velocity, where
\begin{eqnarray}
\label{signal3}
\tan \thetag
= \frac{\bmath{v}_g \cdot \hat{\phi}\times \bmath{v}_p }{\bmath{v}_g
\cdot \bmath{v}_p} 
=\frac{{\cal P} \sin \thetak + {\cal P}'\cos \thetak }{{\cal P} \cos
\thetak - {\cal P}'\sin \thetak } \,. 
\end{eqnarray}
The envelope of the family of such disturbances (whose trajectories are
defined by the angle $\thetak$) constitutes
the Mach cone of fast-magnetosonic waves at any given point in the
flow. It is given by combining equation~(\ref{signal3}) and the condition
\begin{eqnarray}
\label{envelope}
\frac{d}{d\thetak}
\left(\frac{{\cal P} \sin \thetak + {\cal P}'\cos \thetak }{{\cal P}
\cos \thetak - {\cal P}'\sin \thetak } \right) = 0 \,. 
\end{eqnarray}
After some manipulation, equation~\ref{envelope} yields ${\cal P}=0$.\footnote{
The other solution of equation~(\ref{envelope}),
${\cal P}^{''}+{\cal P}=0$, is related to the propagation of
slow-magnetosonic waves; see \cite{Vthesis}.
}
Thus, the fast Mach cone corresponds to a particular combination
($\tilde \thetak\, , \, \tilde \thetag$) of the angles $\thetak$ and
$\thetag$ that satisfies $\tan \tilde \thetag = -\cot \tilde \thetak$
and $ {\cal P}(\tilde \thetak)=0$.  Using equation~(\ref{dispersion}),
these conditions yield
\begin{eqnarray}
\sin^2 \tilde \thetag  =
\nonumber \\
\frac{\displaystyle \frac{B^2 - E^2}{4 \pi w/c^2}+\frac{ c_s^2 } {1-c_s^2/c^2 }
\left[1-  \frac{B_p^2/u_p^2}{4 \pi w /c^2}
\left(1-\frac{r^2}{r^2_{lc}}\right) \right]
}{u_p^2}
\; .
\label{mach_opening}
\end{eqnarray}

The requirement that the Mach cone points outward at a surface $\eta=$const is
that the angle $\arcsin |\hat{\eta}\cdot \bmath{v}_p/v_p |$
between the surface and the poloidal flow velocity exceeds $\tilde \thetag$, or
$ |\hat{\eta}\cdot \bmath{v}_p/v_p |^2 > \sin^2 \tilde \thetag$.
Using equation~(\ref{mach_opening}) and $B_p/v_p=B^{\hat\eta}/v^{\hat\eta}$,
this last inequality can be transformed into equation~(\ref{ext_criterion}).


\end{document}